\documentclass[aps,prl,twocolumn,amsmath,amssymb,showpacs,superscriptaddress,notitlepage,longbibliography,floatfix]{revtex4-1}

\usepackage[colorlinks=true,linkcolor=blue,anchorcolor=red,citecolor=blue, urlcolor=blue]{hyperref}
\usepackage{bm}
\usepackage{graphicx}
\usepackage{xcolor}
\usepackage{cases}

\begin{document}

\title{Signatures of nodal superconductivity in stoichiometric FeTe}

\author{Cequn Li}
\email{cl2775@cornell.edu}
\affiliation{Laboratory of Atomic and Solid-State Physics, Cornell University, Ithaca, NY 14853, USA}

\author{Zi-Jie Yan}
\affiliation{Department of Physics, The Pennsylvania State University, University Park, PA 16802, USA}

\author{Yang Ge}
\affiliation{Department of Physics, University of Florida, Gainesville, FL 32611, USA}
\affiliation{Quantum Theory Project, University of Florida, Gainesville, FL 32611, USA}

\author{Zihao Wang}
\affiliation{Department of Physics, The Pennsylvania State University, University Park, PA 16802, USA}

\author{Bing Xia}
\affiliation{Department of Physics, The Pennsylvania State University, University Park, PA 16802, USA}

\author{Stephen Paolini}
\affiliation{Department of Physics, The Pennsylvania State University, University Park, PA 16802, USA}

\author{Pu Xiao}
\affiliation{Department of Physics, The Pennsylvania State University, University Park, PA 16802, USA}

\author{Lok-Kan Lai}
\affiliation{Department of Physics, The Pennsylvania State University, University Park, PA 16802, USA}

\author{Jiatao Song}
\affiliation{Department of Physics, The Pennsylvania State University, University Park, PA 16802, USA}

\author{Austin R. Kaczmarek}
\affiliation{Laboratory of Atomic and Solid-State Physics, Cornell University, Ithaca, NY 14853, USA}

\author{Lujin Min}
\affiliation{Department of Applied and Engineering Physics, Cornell University, Ithaca, NY 14853, USA}

\author{Kenji Yasuda}
\affiliation{Department of Applied and Engineering Physics, Cornell University, Ithaca, NY 14853, USA}

\author{Peter J. Hirschfeld}
\affiliation{Department of Physics, University of Florida, Gainesville, FL 32611, USA}

\author{Jiabin Yu}
\affiliation{Department of Physics, University of Florida, Gainesville, FL 32611, USA}
\affiliation{Quantum Theory Project, University of Florida, Gainesville, FL 32611, USA}

\author{Cui-Zu Chang}
\email{cxc955@psu.edu}
\affiliation{Department of Physics, The Pennsylvania State University, University Park, PA 16802, USA}

\author{Katja C. Nowack}
\email{kcn34@cornell.edu}
\affiliation{Laboratory of Atomic and Solid-State Physics, Cornell University, Ithaca, NY 14853, USA}
\affiliation{Kavli Institute at Cornell for Nanoscale Science, Ithaca, NY 14853, USA}

\begin{abstract}
Superconductivity in stoichiometric FeTe opens access to the FeTe endpoint of the Fe(Se,Te) phase diagram, yet the nature of its superconducting pairing state remains unresolved. In this work, we combine scanning superconducting quantum interference device (SQUID) microscopy, electrical transport, scanning tunneling microscopy and spectroscopy (STM/S), and mean-field calculations to investigate the local superfluid response and pairing state of FeTe thin films with tunable stoichiometry. Even in stoichiometric FeTe, we observe micrometer-scale spatial variations in both superfluid stiffness and superconducting transition temperature $T_c$, while the London penetration depth remains non‑saturating down to 0.02$T_c$ and follows a power‑law temperature dependence with an exponent of approximately 1--1.5. Together with a V-shaped low-energy density of states and two-gap modeling, these results indicate a superconducting state with gap nodes or deep minima, consistent with either a $d$-wave or nodal $s$-wave superconducting state.
Our findings establish stoichiometric FeTe as a distinct superconducting regime that departs from the trend toward more isotropic gaps at intermediate Se/Te compositions, providing a new benchmark for modern microscopic theories of iron-chalcogenide superconductivity. Our work also reveals a crossover from weak to rapid suppression of $T_c$ as superfluid stiffness decreases, connecting FeTe to the broader phenomenology observed in unconventional superconductors.
\end{abstract}

\maketitle

\section{Introduction}

The discovery of iron chalcogenide superconductors (SCs) has provided a versatile platform for exploring unconventional superconductivity \cite{song2011direct,kasahara2014field,li2016superfluid,sprau2017discovery,liu2018orbital,liang2025pure} and topological superconducting phenomena \cite{wang2018evidence,zhang2018observation,Lin2026}. Within this family, FeSe and FeTe form two contrasting endpoints. FeSe is superconducting and exhibits a highly anisotropic superconducting gap, with evidence supporting either nodal superconductivity \cite{song2011direct,kasahara2014field} or a strongly anisotropic yet finite gap \cite{li2016superfluid,liu2018orbital,liang2025pure,sprau2017discovery}. With increasing Te substitution, the superconducting gap becomes progressively less anisotropic \cite{miao2012isotropic,zhang2018observation,liang2025pure}, suggesting an evolution toward a more isotropic superconducting state as the FeTe endpoint is approached. However, this evolution could not be followed to FeTe itself, which has long been regarded as an antiferromagnetic (AFM) metal with a bi-collinear magnetic ground state that suppresses superconductivity \cite{ma2009firstprinciples,Rodriguez2011,Ducatman2014}. Nonetheless, superconductivity was induced in FeTe through strain \cite{han2010superconductivity} and interface engineering \cite{he2014two,liang2020studies,yao2021hybrid,yi2023dirac,yi2024interface,yi2025universal,sato2025superconductivity,yan2026meissner}, indicating that its nonsuperconducting ground state can be destabilized under appropriately tuned conditions.

This picture has recently changed with the realization of truly stoichiometric FeTe and the discovery of its intrinsic superconductivity~\cite{yan2026stoichiometric,xu2026reversible}. Even nominally stoichiometric FeTe often contains residual interstitial Fe, which stabilizes the AFM order and suppresses Cooper pairing. Post-growth Te-flux annealing removes this interstitial Fe and restores the ideal 1:1 stoichiometry, producing superconductivity with a $T_{\mathrm{c}}\approx12$ K \cite{yan2026stoichiometric}. Therefore, truly stoichiometric FeTe provides access to the superconducting FeTe endpoint of the Fe(Se,Te) phase diagram and raises a central question: does its superconducting state continue the trend toward an increasingly isotropic gap inferred from intermediate Fe(Se,Te) compositions, or does a distinct pairing state emerge at the FeTe endpoint?

\begin{figure}[tbph]
\centering \includegraphics[width=0.45\textwidth]{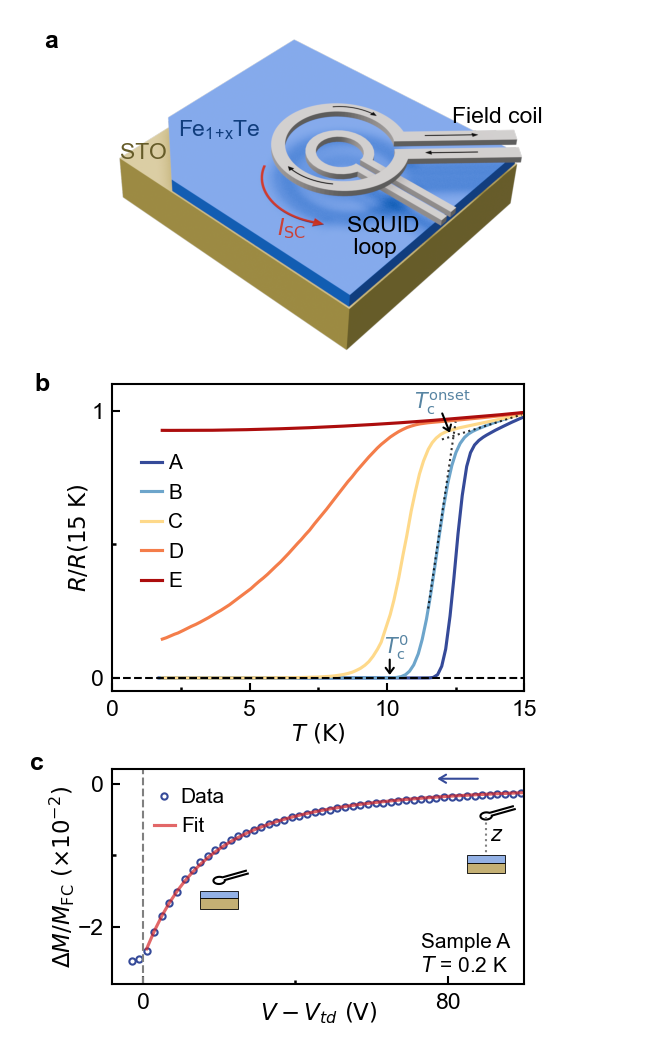}
\caption{\textbf{Diamagnetic response and transport in FeTe.} \textbf{a,} Schematic of the SSM measurement. An ac current through the field coil (inner/outer radii 1.5/3 \textmu m) induces screening currents $I_{SC}$ in $\text{Fe}_{1+x}\text{Te}$, detected by the SQUID loop (inner/outer radii 0.4/1 \textmu m) at a SQUID-sample distance $z\approx1\ \mu m$. The mutual inductance between SQUID and field coil is $M_{\mathrm{FC}}=246\pm1~\mathrm{\phi_0/A}$. \textbf{b,} Normalized longitudinal resistance $R/R$(15~K) versus temperature for samples A–E with interstitial Fe contents $x\approx0$, 0.008, 0.012, 0.015, 0.06, respectively \cite{yan2026stoichiometric}. $T_c^{\mathrm{onset}}$ is defined by the intersection of the extrapolated normal-state resistance and the tangent to the steepest part of the transition, as illustrated for sample B. \textbf{c,} Diamagnetic response $\Delta M/M_{\mathrm{FC}}$ measured as the SQUID approaches sample A. The SQUID-sample distance $z=\beta(V-V_{td})+z_0$, where $V$ is the piezo voltage, $\beta$ the bender constant, and $V_{td}$ and $z_0$ are the voltage and SQUID–sample distance at touchdown, respectively. The best fit yields $\beta=63\pm3$ nm/V and $z_0=1.5\pm0.3\ \mathrm{\mu m}$ (see Supplementary Section S12).}
\label{fig:setup}
\end{figure}

Here we address this question by imaging the local superfluid stiffness $K_s$ of FeTe thin films using scanning SQUID microscopy (SSM). Within London theory, $K_s=(\hbar^2/4)\sum_i n_{s,i}/m_i^\ast \propto \lambda^{-2}$ for a multiband SC, where $\hbar$ is the reduced Planck constant, $n_{s,i}$ and $m_i^\ast$ are the superfluid density and effective mass of band $i$, respectively, and $\lambda$ is the London penetration depth~\cite{tinkham2004introduction,Huang2016theoretical}. Therefore, the temperature dependence $\lambda(T)$ reflects the depletion of the superfluid density by thermally excited quasiparticles. A fully gapped SC exhibits an exponential low-temperature dependence, whereas a nodal order parameter leads to a power‑law variation of $\lambda(T)$ \cite{prozorov2006magnetic,hicks2009evidence,fletcher2009evidence,hashimoto2010line,ferguson2024local}. In stoichiometric FeTe, we find that $\lambda$ remains non-saturating down to $0.02T_c$ and follows a power-law temperature dependence with an exponent of $\sim1$--1.5. STM/S independently reveals a V-shaped low-energy density of states (DOS), further supporting a superconducting gap with nodes or deep minima. Together with mean-field calculations, these results point to a two-gap superconducting state in which one gap is strongly anisotropic or nodal, challenging existing expectations for the conventional Fe(Se,Te) phase diagram. We further observe micrometer-scale variations in $K_s$ and $T_c$, as well as a crossover from a weak dependence of $T_c$ on $K_s$ at large superfluid stiffness to a rapid suppression of $T_c$ as $K_s$ approaches the scale of $T_c$, showing some similarities to the broader Uemura phenomenology \cite{Uemura1989} and to trends reported in overdoped cuprates \cite{bovzovic2016dependence} and infinite-layer nickelates \cite{li2026correlation}. These results establish stoichiometric FeTe as a distinct regime of iron-based superconductivity with an unconventional pairing state.

\section{Spatial inhomogeneity of superfluid stiffness and transition temperature}

\begin{figure*}[tbph]
\centering \includegraphics[width=.7\textwidth]{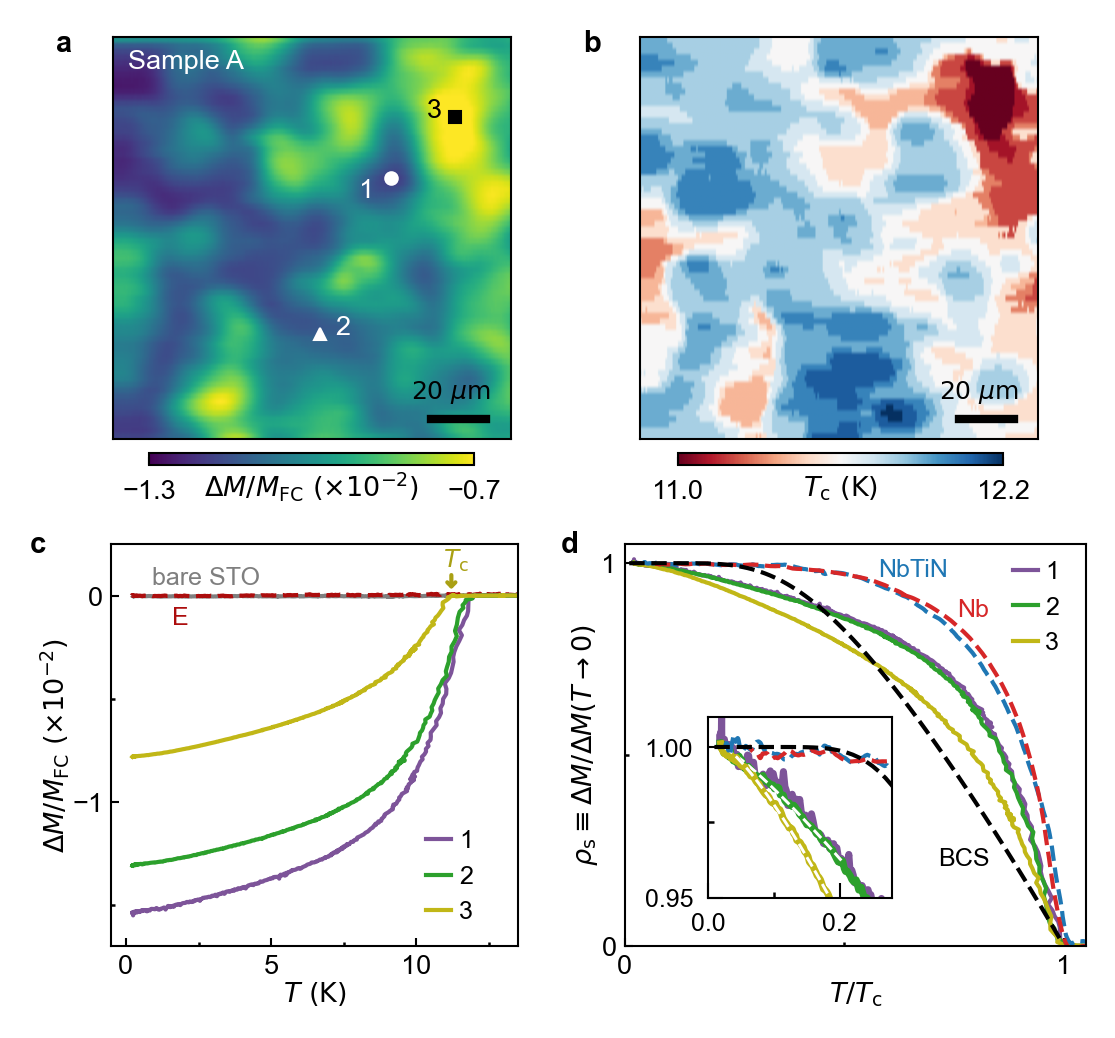}
\caption{\textbf{Spatial variation of superconductivity and non-saturating superfluid stiffness.} \textbf{a,} Spatial map of the diamagnetic response $\Delta M/M_{\mathrm{FC}}$ in sample A at $T=0.2$ K at a constant height, $V-V_{td}=15$ V. \textbf{b,} Spatial map of $T_c$ over the same field of view. $T_c$ is defined as the temperature at which $\Delta M/M_{\mathrm{FC}}$ reaches a threshold of $5\times10^{-5}$ at each pixel. \textbf{c,} Temperature dependence of $\Delta M/M_{\mathrm{FC}}$ measured at selected locations indicated in \textbf{a}, at $V-V_{td}=12\ \mathrm{V}$. Bare STO (gray) and sample E (red) show no discernible response. \textbf{d,} Normalized superfluid stiffness $\rho_s$ versus $T/T_c$. Blue and red dashed lines are data measured on NbTiN ($d=25$ nm) and Nb ($d=20$ nm) thin films, respectively; the black dashed line is the clean-limit BCS prediction for $\Delta=1.76k_BT_c$. The inset shows a zoomed-in view near $T=0$. White dashed lines are power-law fits, $\rho_{s}=1-b(T/T_c)^n$, for $T\leq0.2\ T_c$. For the green and purple curves, $b=0.29\pm0.02$ and $n=1.18\pm0.03$; for the yellow curve, $b=0.52\pm0.02$ and $n=1.29\pm0.02$.}
\label{fig:susceptibility}
\end{figure*}

 Figure~\ref{fig:setup}(a) illustrates our experimental setup. The SQUID susceptometer consists of two concentric coils: an inner pickup loop, which forms part of the SQUID circuit and detects magnetic flux, and an outer field coil, which generates a local AC magnetic field  \cite{huber2008gradiometric}. Far from the FeTe sample, the mutual inductance between the two coils, $M_{\mathrm{FC}}$, is set by their geometry. As the coils approach the sample, the local magnetic field induces Meissner screening currents in FeTe. These currents create a compensating flux through the pickup loop that results in a decrease in the measured mutual inductance, $\Delta M$. A ratio $\Delta M/M_{\mathrm{FC}}=-1$ corresponds to complete screening of the magnetic field in the plane of the SQUID loop.

In the Pearl limit $\lambda\gg d$ where $d$ is the film thickness, the absolute two-dimensional (2D) superfluid stiffness, expressed in temperature units, is $K_s=\upsilon d/(2\lambda^2)$, where $\upsilon=\hbar^2/(2\mu_0k_Be^2)\approx1.24~\mathrm{K\cdot cm}$, $\mu_0$ is the magnetic permeability of vacuum, $k_B$ is the Boltzmann constant, and $e$ is the electron charge~\cite{bovzovic2016dependence,zhang2026imaging}. The measured response can be directly related to $K_s$ through $\Delta M / M_\mathrm{FC} = - \frac{L_{geo}}{\upsilon}K_s$ where $L_{geo}$ is a geometric length scale determined by the sensor geometry and the SQUID-sample distance $z$, approaching the field coil radius as $z\rightarrow 0$ \cite{kirtley2012scanning}. At fixed $z$, this geometric factor cancels upon normalization to the zero-temperature response: $\Delta M(T)/\Delta M(0)=\lambda^2(0)/\lambda^2(T)=K_s(T)/K_s(0)=\rho_s(T)$ where $\rho_s(T)$ is the normalized superfluid stiffness. Experimentally, we approximate the zero-temperature values by those measured at our base temperature, $T=0.2~\mathrm{K}$. The independence of the normalized response from $z$ is confirmed by the collapse of $\Delta M(T)/\Delta M(0)$ curves obtained at different $z$ onto a single trace (see Supplementary Fig. S6). The normalized superfluid stiffness and related quantities can therefore be obtained directly from the measured signal \cite{kirtley2012scanning,Bert2012,ferguson2024local}.

Multiple FeTe thin films grown by molecular-beam epitaxy (MBE) on SrTiO$_3$ (STO) substrates are examined in this work, each with a nominal thickness of approximately 40 unit cells ($\sim$ 25 nm). The interstitial Fe content in these films is controlled by varying the duration of the Te‑annealing cycles. A full description and characterization of the growth is given in Ref. \cite{yan2026stoichiometric}. Sample A is annealed under the optimal conditions identified in Ref. \cite{yan2026stoichiometric} and exhibits zero longitudinal resistance below $T_c^0=12$ K as shown in Fig. \ref{fig:setup}(b), consistent with the stoichiometric samples reported there. From samples A to D, the Fe:Te ratio progressively increases, leading to a systematic suppression of $T_c$ accompanied by a broadening of the superconducting transition. Sample E is non‑superconducting as‑grown FeTe and used as a control sample.

Figure \ref{fig:setup}(c) shows an example of the diamagnetic response measured as the SQUID approaches the surface of sample A at $T=0.2$ K, plotted as a function of the voltage applied to the $z$-piezo. Using the model from Ref. \cite{kirtley2012scanning}, we extract $\lambda(0)=522\pm22$ nm at the location of this approach curve, similar to values reported for FeSe \cite{kasahara2014field} and FeSe$_{1-x}$Te$_x$ \cite{liang2025pure}. As shown below, the superfluid stiffness, and hence the London penetration depth, varies spatially across the sample. Details of the absolute $\lambda(0)$ fitting procedure are provided in Supplementary Section S12. The uncertainty in $\lambda(0)$ is dominated by uncertainties in the $z$-piezo calibration and the SQUID–sample height offset at touchdown, arising from the finite setback of the SQUID from the chip tip and the chip–sample alignment angle (see Supplementary Fig. S1). It does not account for additional systematic uncertainty associated with the simplified SQUID geometry assumed in the model, which is difficult to quantify and may increase the overall uncertainty in $\lambda(0)$.

Figure \ref{fig:susceptibility}(a) presents the spatially resolved diamagnetic response in sample A at $T=0.2$ K under zero background magnetic field (see field compensation and image of vortices in Supplementary Fig. S2). The response is spatially inhomogeneous, revealing nearly twofold variations in the superfluid stiffness across the field of view, with darker regions corresponding to stronger superfluid stiffness. Autocorrelation analysis yields a characteristic length scale of approximately 10--20 $\mathrm{\mu m}$ for these spatial variations (see Supplementary Fig. S4), larger than the spatial resolution of our SQUID of $\sim$3--6 $\mathrm{\mu m}$ determined by the field-coil size. No spatial periodicity or long‑range ordering is observed. Similar spatial variations in the Meissner response were observed using magnetic force microscopy in stoichiometric FeTe \cite{yan2026stoichiometric} and FeTe-based heterostructures \cite{yan2026meissner}. Figure \ref{fig:susceptibility}(c) shows  $\Delta M(T)/M_\mathrm{FC}$ traces obtained at locations with the strongest, weakest, and intermediate diamagnetic response across the field of view. We define $T_c$ as the temperature at which the diamagnetic response first becomes detectable upon cooling.

\begin{figure*}[tbph]
\centering \includegraphics[width=0.75\textwidth]{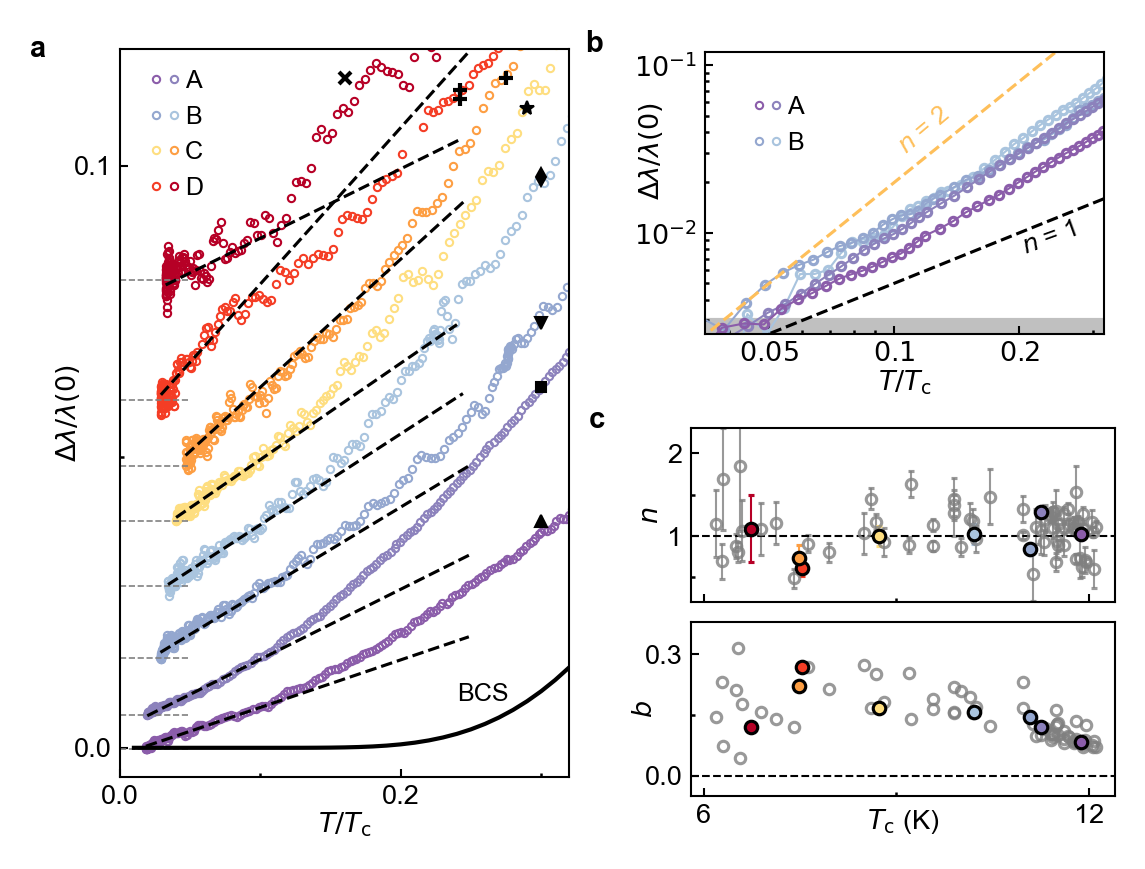}
\caption{\textbf{Power-law temperature dependence of penetration depth.} \textbf{a,} Temperature dependence of the normalized change in the London penetration depth, $\Delta\lambda/\lambda(0)$, at selected locations in samples A--D. Symbols indicate the corresponding locations in Fig. \ref{fig:susceptibility}\textbf{a} and Figs. \ref{fig:correlation}\textbf{a}--\textbf{c}. Curves are offset for clarity, with offsets indicated by horizontal gray dashed lines. Black dashed lines are linear fits for $T<0.12T_c$. \textbf{b,} Data from \textbf{a} for samples A and B on a log-log scale. Black and yellow dashed lines represent linear and quadratic scaling, respectively; gray shading indicates the noise level. \textbf{c,} Exponent $n$ (top) extracted from power-law fits and slope $b$ (bottom) extracted from linear fits for $T<0.12T_c$ versus  local $T_c$. The error bars are smaller than the dots. Color-coded points correspond to data in \textbf{a} and open gray points to additional data (Supplementary Fig. S10).}
\label{fig:lambda}
\end{figure*}

Figure \ref{fig:susceptibility}(b) shows the spatial distribution of $T_c$ in the same field of view, extracted from a series of images acquired over a range of temperatures (see representative images in Supplementary Section S3). The spatial structure of the $T_c$ map resembles the low-temperature diamagnetic response image, with a peak-to-peak variation $\Delta T_c\approx1.2$ K. This variation is consistent with the transition width measured in transport, $T_c^{\mathrm{onset}}-T_c^0$ (Fig. \ref{fig:setup}(b)). The median local $T_c$ approximately coincides with the temperature at which the sample resistance falls to 20\% of its normal-state resistance, $R$(15 K) (see Supplementary Fig. S4). Similar spatial variations are observed in samples A--D and become progressively more pronounced from A to D, as shown in Figs. \ref{fig:correlation}(a)--(f). Across these samples, the local $T_c$ distributions remain closely connected to the macroscopic transport properties. Consistent with the absence of a resistive transition, no discernible diamagnetic response is observed in sample E as well as bare STO.

The physical origin of the micrometer-scale variations in $K_s$ and $T_c$ is currently unknown. We examined the surface morphology of sample A using atomic force microscopy but found no correlation between the topography and the local $K_s$ (see Supplementary Fig. S5). The characteristic lateral size of the surface features is $\sim0.6\ \mathrm{\mu m}$, much smaller than the length scale discussed above. Other possibilities include structural defects, strain, or nonuniform distribution of residual interstitial Fe both in the lateral directions and through the film thickness. Further studies are required to clarify this aspect. Regardless of their origin, the close similarity between the diamagnetic-response map and the $T_c$ map indicates a correlation between local $K_s$ and $T_c$, which we discuss below.

\section{Power-law temperature dependence of London penetration depth}

Strikingly, the $\Delta M(T)/M_\mathrm{FC}$ curves in Fig. \ref{fig:susceptibility}(c) continue to vary down to 0.02 $T_c$, without any indication of saturation. We performed control measurements to rule out possible experimental artifacts, including nonlinear response generation, thermal lag, and piezo drift (see Supplementary Section S7). Figure \ref{fig:susceptibility}(d) shows $\rho_{s}$ as a function of $T/T_c$. For an isotropic, fully gapped Bardeen–Cooper–Schrieffer (BCS) SC, $\rho_{s}$ is expected to saturate below $\sim$0.25 $T_c$ as the gap becomes fully developed \cite{prozorov2006magnetic}. We observe this expected behavior in conventional thin-film SCs, such as Nb \cite{lemberger2007penetration} and NbTiN \cite{hong2013terahertz,PerezLozano2024} measured using the same setup. In contrast, below 0.2 $T_c$, the $\rho_{s}(T)$ curves measured at three locations selected to span the range of diamagnetic response in sample A follow a power-law dependence, $(1-\rho_{s})\propto T^n$  with an average exponent $n=1.2\pm0.1$. Such low-temperature behavior is usually associated with a gap structure hosting either nodes or deep minima \cite{prozorov2006magnetic,hicks2009evidence,fletcher2009evidence}.

To examine the low-temperature power-law behavior more closely across samples A--D, we express the data in terms of the normalized change in London penetration depth, $\Delta\lambda/\lambda(0)=\lambda(T)/\lambda(0)-1$, as shown in Fig. \ref{fig:lambda}(a). Power-law scaling is evident from the log-log plot in Fig. \ref{fig:lambda}(b), which yields an overall slope corresponding to $n\approx1.5$ with a tendency toward linear scaling near zero temperature. Power-law fits to the data for $T/T_c\leq0.12$ give a consistent exponent $n\approx1$, nearly independent of the local $T_c$ (see top panel of Fig. \ref{fig:lambda}(c)). Supplementary Fig. S11 shows the fitted $n$ extracted over different temperature ranges. We also impose linear fits of the form $\Delta\lambda/\lambda(0)=b(T/T_c)$ for $T/T_c\leq0.12$, which describe the data reasonably well. The resulting slope $b$ remains nonzero, in clear contrast to conventional BCS behavior.

For nodal SCs, the exponent depends on the dimensionality of nodes and the disorder level \cite{prozorov2006magnetic}. The observed values $n \approx 1$--1.5 are consistent with a gap with line nodes in the presence of weak- to moderate-strength impurities. As a consistency check, we use the Drude expression and estimate the mean free path $l$ to be on the order of 1 nm, comparable to the superconducting coherence length $\xi\approx2$ nm \cite{xu2026reversible,yan2026stoichiometric}, placing stoichiometric FeTe between the clean and dirty limits (see details in Supplementary Section S9). Alternatively, the observed behavior could arise from a very small BCS gap or gap minima, $\Delta_{min}$, which is difficult to distinguish from true nodes given the finite measurement resolution and base temperature. From exponential fits to $\Delta\lambda/\lambda(0)$ at low temperature, we obtain a conservative upper bound for $\Delta_{min}$ of 0.1 meV, corresponding to roughly $0.1$ $k_BT_c$ (see Supplementary Section S10).

\begin{figure*}[tbph]
\centering \includegraphics[width=1\textwidth]{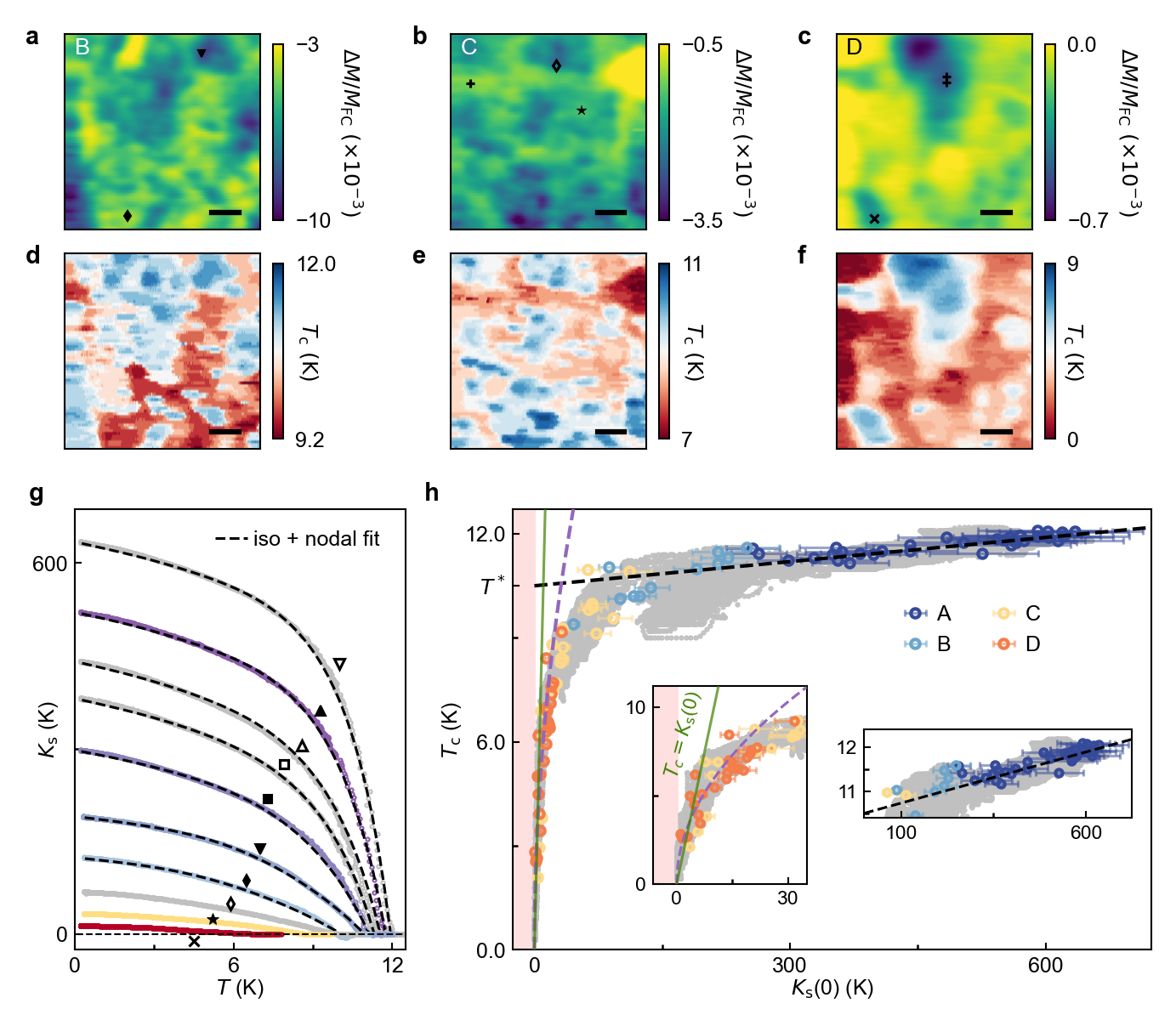}
\caption{\textbf{Scaling between transition temperature and superfluid stiffness.} \textbf{a--c,} Diamagnetic-response maps at $T=0.2$ K and $V-V_{td}=15\ \mathrm{V}$ in samples B--D, respectively. \textbf{d--f,} Corresponding $T_c$ maps. Scale bar, 20 $\mathrm{\mu m}$. \textbf{g,} Temperature dependence of $K_s$ at selected locations in samples A--D, indicated in \textbf{a}--\textbf{c}, Fig. \ref{fig:susceptibility}\textbf{a}, and Supplementary Figs. S10 and S14. Color-coded curves correspond to the data in Fig. \ref{fig:lambda}\textbf{a}. Black dashed lines denote the two-gap fits. \textbf{h,} Local $T_c$ versus $K_s(0)$. Gray points are extracted from the spatial maps, while color-coded points are obtained from approach curves and temperature sweeps at selected locations. Insets highlight the high- (right) and low-stiffness (left) regimes. The black dashed line is $T_c=0.0023K_s(0)+10.5\ \mathrm{K}$, and the purple dashed line a square-root dependence adopted from Refs. \cite{bovzovic2016dependence,li2026correlation}, $T_c=(1.886\ \mathrm{K}^{1/2}) \sqrt{K_s(0)}$. The crossover occurs near $T^*=10.5$ K. The green solid line corresponds to $T_c=K_s(0)$.}
\label{fig:correlation}
\end{figure*}

Another mechanism that can produce near-power-law $\Delta\lambda(T)$ without  nodes in the gap is phase fluctuations (PFs) between weakly coupled, fully gapped superconducting regions in strongly disordered  SCs \cite{lamura2002granularity,khvalyuk2024near}. However, the reported effect is typically much weaker than that observed in stoichiometric FeTe: the low-temperature slope $b$ in $\Delta\lambda/\lambda(0)=b(T/T_c)$ for disordered NbN \cite{lamura2002granularity} and amorphous InO$_x$ thin films \cite{khvalyuk2024near} is at least an order of magnitude smaller. For thermal PFs associated with disorder-induced low-energy collective modes, the magnitude of the effect depends on the ratio $T_c/K_s(0)$~\cite{khvalyuk2024near}. When $T_c/K_s(0)\ll1$, PF-induced corrections become negligible and the thermally activated BCS behavior is recovered. To assess this scenario quantitatively, we compare the local $T_c$ with $K_s(0)$.
The extracted $K_s(0)$ decreases monotonically with decreasing $T_c$ (Figs. \ref{fig:correlation}(g) and (h)). For the stoichiometric sample, $K_s(0)$ reaches $\sim630$~K, corresponding to a 2D superfluid density $n_s\approx3\times10^{15}\ {\rm cm^{-2}}$ estimated assuming a single band with $m^*\approx10m_e$ \cite{Lin2026}, comparable to the normal-state carrier density $n_{2D}\approx10^{16}\ \mathrm{cm}^{-2}$ obtained from Hall measurements in sample A. Because $K_s(0)\gg T_c$ in this regime, PFs are unlikely to account for the pronounced non-saturating $\lambda(T)$. Consistently, a phenomenological PF model \cite{khvalyuk2024near} cannot reproduce the observed temperature dependence for the small ratio $T_c/K_s(0)$ in the $T_c\approx12$~K limit (see comparisons in Supplementary Section S13). At lower $T_c$, however, $T_c/K_s(0)$ increases rapidly, and PFs may become more important. Indeed, in the lower-$T_c$ samples with pronounced spatial inhomogeneities, the slope $b$ extracted from linear fits in Fig. \ref{fig:lambda}(a) for $T\leq0.12\ T_c$ develops substantial scatter (bottom panel of Fig. \ref{fig:lambda}(c)). Consistent with this trend, $\Delta\lambda(T)$ exhibits pronounced spatial variation and frequently displays double transitions across sample D, likely reflecting the SQUID response averaged over multiple superconducting regions (see additional data in Supplementary Fig. S10). These observations indicate that inhomogeneity becomes increasingly important as $T_c$ is suppressed. We therefore focus on the $T_c\approx12$ K regime when drawing conclusions about the gap symmetry.

\begin{figure*}
\centering \includegraphics[width=0.75\textwidth]{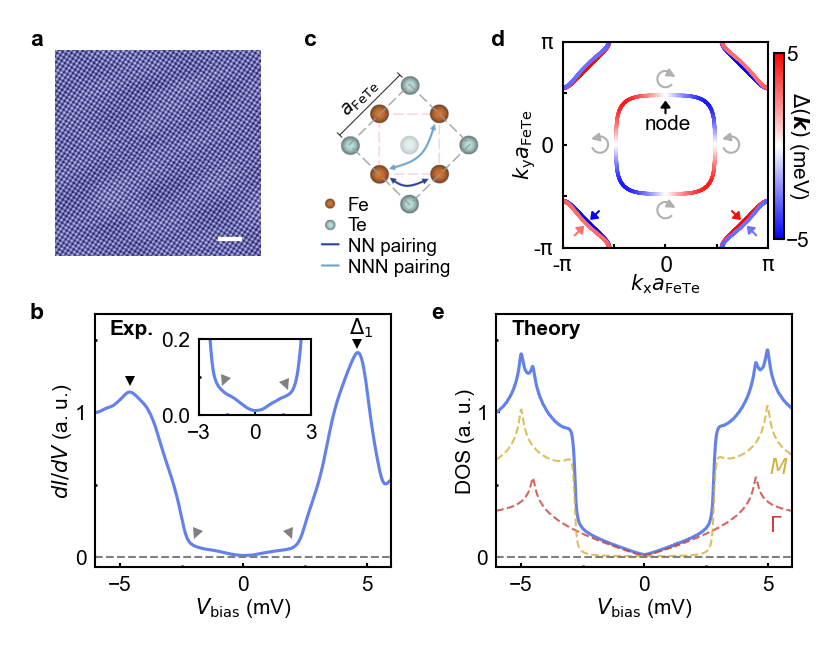}
\caption{\textbf{Evidence for nodal multigap superconductivity from spectroscopy and modeling.} \textbf{a,} STM image (20$\times$20 nm$^2$) of a 40-unit-cell stoichiometric FeTe film similar to sample A. Scale bar, 2 nm. \textbf{b,} Corresponding $dI/dV$ spectrum at $T=0.31$ K. A larger‑scale view is shown in Supplementary Fig. S16. \textbf{c,} Top view of the FeTe lattice. Dark and light blue lines indicate attractive interactions between the NN and NNN Fe sites included in the model.
\textbf{d,} The superconducting gap $\Delta(\boldsymbol{k})$ on the renormalized Fermi surfaces  obtained from the mean-field calculation.
The pocket centered at $\Gamma$ $(0,0)$ is nodal with $d$-wave symmetry and an approximately $\sin(2\varphi)$ angular dependence. The nodes are identified by their chiral winding numbers, indicated by clockwise ($-1$) and anticlockwise ($+1$) light-gray arrows. The two symmetry-related, closely spaced Fermi pockets centered at $M(\pm \pi,\pm \pi)$ are fully gapped with opposite pairing signs (red and blue arrows) and gap magnitudes ranging from $\sim3$~meV to $\sim5$~meV (see Supplementary Fig. S19).
\textbf{e,} Calculated local DOS at $T=0$. The gap nodes produce a linear V-shaped low-energy DOS within the broader gap. Red and yellow dashed lines denote contributions from the $\Gamma$ and $M$ pockets, respectively.
}
\label{fig:gap-structure}
\end{figure*}

\section{Signatures of nodal two-gap superconductivity}

To gain insight into the gap symmetry, we fit the $K_s(T)$ curves in Fig. \ref{fig:correlation}(g) using several gap models. Single‑gap models fail to describe the data over the full temperature range (see details in Supplementary Section S11), motivating adoption of a simple two-gap model, i.e., $K_s\propto\rho_s^{total}=(1-r)\rho_{s1}+r\rho_{s2}$, where $0\leq r\leq1$ is the weighting factor and $\rho_{s1}$ and $\rho_{s2}$ are the normalized superfluid stiffnesses associated with gaps $\Delta_{1}$ and $\Delta_{2}$, respectively \cite{johnston2013elaboration,li2016superfluid}. We assume a common $T_c$ for both gaps because no secondary transition or kink is observed in the measured $K_s(T)$. We first consider two isotropic gaps; however, the corresponding fit predicts a saturating $\rho_s^{\mathrm{total}}$ below $T\sim0.1T_c$, in disagreement with the observed behavior. We therefore consider a nodal form for the second gap, $\Delta_2(\varphi)=\Delta_2\sin(2\varphi)$ ($\varphi$ is the angle along the Fermi surface) \cite{sprau2017discovery,liu2018orbital,liang2025pure}, while retaining an isotropic form for $\Delta_1$. As shown in Fig. \ref{fig:correlation}(g), the resulting fits capture the data well. The three fit parameters are strongly correlated and therefore cannot be independently constrained by the superfluid stiffness alone. To assess the range of gap values consistent with the data, we fix $r$ and fit $\Delta_1$ and $\Delta_2$ for different values of $r$. For the $T_c=12$~K data, fits with essentially indistinguishable quality are obtained over ranges $\Delta_1=$3.2–4~meV and $\Delta_2=$1.6–5~meV  (see Supplementary Section S11).

The two-gap scenario is further corroborated by the tunneling spectrum obtained from STM/S measurements. Figure \ref{fig:gap-structure}(a) shows a defect-free 20$\times$20-nm$^2$ STM topograph of a stoichiometric FeTe film, similar to sample A, confirming the absence of interstitial Fe \cite{yan2026stoichiometric}. The corresponding tunneling spectrum, shown in Fig. \ref{fig:gap-structure}(b), exhibits a pronounced U-shape with coherence peaks at bias voltage $V_\mathrm{bias}\approx 4.6$ mV, yielding a ratio $\Delta_{1}/k_BT_c\approx 4$, substantially exceeding the weak-coupling BCS value of 1.764~\cite{tinkham2004introduction}, but comparable to maximum-gap ratios of $\sim$ 3.6 commonly observed in Fe-based SCs~\cite{Miao2018}. A full characterization of this gap, including its temperature dependence and disorder-induced in-gap states, can be found in Ref. \cite{yan2026stoichiometric}.
Strikingly, however, the spectrum does not develop a flat bottom within this gap. Instead, following a steep decrease from the coherence peaks, the $dI/dV$ crosses over near $V_\mathrm{bias}\approx\pm2$~mV to a V-shaped bias dependence. The overall shape therefore points to multiple superconducting energy scales, including a low-energy or nodal gap giving rise to a V-shaped DOS. This observed low-energy DOS contrasts with the U-shaped gap reported in FeTe$_{0.55}$Se$_{0.45}$ \cite{wang2018evidence}, and together with the nearly linear $\Delta\lambda(T)$, provides further evidence for a nodal gap in stoichiometric FeTe.

Theoretical models of superconductivity in FeTe have previously considered nonlocal pairing between next-nearest-neighbor (NNN) Fe sites~\cite{gastiasoro2016unconventional,miao2012isotropic} and explored the competition between antiferromagnetism and superconductivity~\cite{yan2026stoichiometric}. NNN pairing can produce zero lines within the Brillouin zone and therefore in principle nodal SC; however, the Fermi pockets around $\Gamma$ and $M$ points in FeTe are expected to be small, as found in density functional theory, such that these zero lines do not intersect the Fermi surfaces and the resulting gap is nodeless. We therefore extend the interacting model for FeTe from Ref.~\cite{yan2026stoichiometric} by adding nearest-neighbor (NN) attractive interactions, another pairing channel previously considered for iron chalcogenides~\cite{miao2012isotropic,hung2012anisotropic}, to the NNN interaction as illustrated in Fig.~\ref{fig:gap-structure}(c). We then explore whether nodal gaps can emerge without substantially altering the Fermi-surface structure or interaction strengths of the original model.

We perform self-consistent mean-field calculations at zero temperature, as described in Methods and Supplementary Section S15. The interactions renormalize the Fermi surface, suppressing the two inner hole pockets while keeping the outer hole pocket around the $\Gamma$ point. Superconducting gaps develop on the remaining Fermi pockets around $\Gamma$ and $M$. Over a broad range of interaction strengths, we find either a $d$-wave state or a nodal $s$-wave state, demonstrating that nodes can emerge without fine tuning of the model parameters. The presence and location of the nodes depend on details of the underlying Fermi surface, which has not yet been experimentally determined.

To reproduce the main features observed in the $dI/dV$ spectrum in Fig.~\ref{fig:gap-structure}(b), including the position of coherence peaks and low-energy spectral features, the parameters are more carefully chosen. Figure~\ref{fig:gap-structure}(d) shows one representative self-consistent solution, for parameters specified in Methods, that yields a spin-singlet $d$-wave state and approximate agreement with the measured DOS. In this solution, the gap on the $\Gamma$ pocket develops nodes, whereas the two symmetry-related $M$ pockets remain fully gapped, with gaps of opposite signs.
The nodes carry nontrivial chiral winding numbers protected by the spinless chiral symmetry of the system (see Methods)~\cite{wen1989winding,volovik2007qpt,sato2011andreev,schnyder2011surface,chiu2016classification}.  The coexistence of these nodal and nodeless gaps naturally produces a two-gap DOS line shape as shown in Fig.~\ref{fig:gap-structure}(e), similar to the observed spectrum. In the calculation, the DOS peaks at $V_\mathrm{bias}\approx5$~mV arise from the gap maxima on the $\Gamma$ and $M$ pockets, while the shoulder at $V_\mathrm{bias}\approx3$~mV reflects the gap minimum on the $M$ pockets, potentially accounting for the $\Delta_1$ peak and the $\sim2$~mV crossover point in the measured spectrum in Fig.~\ref{fig:gap-structure}(b), respectively.

In Supplementary Section~S15, we present additional parameter sets that yield nodal states and comparable DOS line shapes. These include a $d$-wave solution in which the $\sim2$~meV feature instead corresponds to the maximum of the nodal gap on the $\Gamma$ pocket, as well as a nodal $s$-wave solution obtained upon including a next-next-nearest-neighbor (NNNN) attractive interaction. Thus, while the present results do not distinguish between $d$- and $s$-wave states, they establish nodal superconductivity as a robust outcome across different parameter regimes. This is consistent with the robust near-linear $\Delta\lambda(T)$ observed across the stoichiometric sample despite spatial variations in superfluid stiffness and $T_c$.

\section{Correlation between superfluid stiffness and transition temperature}

Our spatially resolved measurements reveal an additional notable feature: a correlation between the local $T_c$ and $K_s(0)$, as shown in Fig.~\ref{fig:correlation}(h). At large $K_s(0)$, $T_c$ depends only weakly on the stiffness, whereas it is rapidly suppressed once $K_s(0)$ approaches the scale of $T_c$. The overall trend resembles the broader Uemura phenomenology in cuprates, where $T_c$ approximately scales with $K_s(0)$ in the underdoped, low-stiffness regime but tends to saturate with increasing stiffness as optimal doping is approached~\cite{Uemura1989}. Similar crossover behavior has also been reported in several other SCs, including overdoped La$_{2-x}$Sr$_x$CuO$_4$ (LSCO) \cite{bovzovic2016dependence}, the infinite-layer nickelate Nd$_{0.8}$Sr$_{0.2}$NiO$_2$ (NSNO) \cite{li2026correlation}, and disorder-tuned amorphous InO$_x$ \cite{Charpentier2025}, although the microscopic circumstances differ substantially. Compared with the LSCO and NSNO data, our measurements extend further into the regime $K_s(0)\gg T_c$, in which we find that $T_c$ varies only weakly with $K_s(0)$. A linear fit for $K_s(0)>150$~K using $T_c=T_c^0+\alpha K_s(0)$ yields $T_c^0=10.5$~K and $\alpha=0.0023$.

The superfluid stiffness sets the energy scale for spatial variations in the superconducting phase, so PFs can become important when $K_s(0)$ approaches the scale of $T_c$ \cite{Emery1995}. We include the line $T_c=K_s(0)$ in Fig.~\ref{fig:correlation}(h) as a reference for the regime in which these energy scales become comparable. All data lie at or below this line, with only the lowest-stiffness points approaching it. Nevertheless, the origins of the weak dependence at large $K_s(0)$, the crossover, and the rapid suppression of $T_c$ at low $K_s(0)$ remain unclear. Although $T_c$ may become limited by PFs at low stiffness, disorder provides another natural explanation. Calculations for disordered nodal SCs show that pair breaking can produce a strongly nonlinear, correlated suppression of $T_c$ and $K_s(0)$, including crossover behavior qualitatively similar to that observed here \cite{franz1997critical,LeeHone2017,Pal2023}. The pronounced spatial variations in the lower-$T_c$ samples directly demonstrate an inhomogeneous regime in which theories based on homogeneous, disorder-induced pair breaking may no longer be sufficient. Such inhomogeneity may produce weakly coupled superconducting regions and thereby enhance PFs \cite{RamshawKivelson2026}. Because $T_c$ depends on the concentration of interstitial Fe \cite{yan2026stoichiometric}, a possible source of the observed inhomogeneity is a nonuniform distribution of interstitial Fe, both laterally and through the film thickness. SSM probes the screening response integrated through the full film thickness and therefore complements surface-sensitive probes such as STM.

\section{Discussion and outlook}

Finally, we place our results in the broader context of the Fe(Se,Te) family. Increasing Te substitution has been associated with a reduction in superconducting gap anisotropy, accompanied by changes in the nematic and orbital-selective electronic structure \cite{liang2025pure}. Our observations in stoichiometric FeTe deviate from this trend, pointing to a distinct superconducting regime at the FeTe endpoint. Therefore, the re-emergence of strong gap anisotropy may reflect a change in the balance among competing pairing channels~\cite{miao2012isotropic,hung2012anisotropic}. The pronounced micrometer-scale spatial variations of $K_s$ and $T_c$ further reveal unexpected superconducting inhomogeneity, whose microscopic origin and possible relationship to the superconducting state remain important questions for future investigation.

In summary, we demonstrate signatures of nodal superconductivity in stoichiometric FeTe thin films through local superfluid-stiffness imaging. The power-law temperature dependence of $\lambda$ with an exponent $n$ of 1--1.5 indicates the presence of gap nodes or deep gap minima with $\Delta_{min}<0.1k_BT_c$, corroborated by the V-shaped low-energy DOS revealed by STM/S. Mean-field calculations suggest that these gap nodes may arise from either $d$-wave or nodal $s$-wave pairing associated with Fe sites. Full microscopic calculations to support the phenomenological model of pairing presented here are in progress. In addition, we establish a correlation between the local $T_{\mathrm{c}}$ and $K_s(0)$, resembling trends reported in high-$T_c$ cuprates and NSNO. Together, these results establish stoichiometric FeTe, one of the simplest iron-based SCs, as a promising platform for investigating the microscopic origin of unconventional superconductivity.

\textit{Note added:} During preparation of this manuscript, we became aware of a theoretical study of superconducting pairing in stoichiometric FeTe~\cite{Hua2026} that finds a nodeless superconducting state at the stoichiometric filling, providing a complementary theoretical perspective on the superconducting gap structure.

\section{Acknowledgements}
C.L., A.R.K., and K.C.N. acknowledge support from the Air Force Research Laboratory, Project Grant FA9550-21-1-0429. The electrical transport measurements were supported by an NSF grant (DMR-2241327). The MBE growth and STM/S measurements were supported by a DOE grant (DE-SC0023113). C.-Z.C. acknowledges the support from the Gordon and Betty Moore Foundation’s EPiQS Initiative (GBMF9063). Y.G. and J.Y.'s work is supported by J.Y.'s startup funds at University of Florida. P.J.H. was supported by NSF-DMR-2231821. L.M. and K.Y. acknowledge support from DOE grant (DE-SC0026110). We thank Ruslan Prozorov, Brad J. Ramshaw, Eun-Ah Kim, and Ke Wang for helpful discussions. We thank Maciej W. Olszewski, Dan Ralph, and Valla Fatemi for providing the Nb films. We thank the imec Superconducting Digital Program team for providing the NbTiN films, which were grown with support from the Army Research Office under Grant No. W911NF-24-1-0150.

\section{Author contributions}
C.L. and K.C.N. conceived the experiment; C.L. performed the scanning SQUID measurements and data analysis with assistance from A.R.K., under the supervision of K.C.N.; Z.-J.Y., P.X., and L.-K.L. performed the MBE growth and transport measurements under the supervision of C.-Z.C.; Z.W., B.X., S.P., and J.S. performed the STM/S measurements under the supervision of C.-Z.C.; L.M. performed the atomic force microscopy measurements under the supervision of K.Y.; Y.G., P.J.H., and J.Y. provided theoretical support; C.L. and K.C.N. wrote the manuscript with input from all authors.

\section{Methods}
\subsection{MBE growth and Te-annealing treatments}
FeTe films used in this work are grown in two commercial MBE chambers [1 Lab 10 from ScientaOmicron and 1 from Unisoku]. Each MBE chamber has a vacuum better than $3\times10^{-10}$ mbar. Both metallic 0.5\% Nb-doped SrTiO$_3$(100) and insulating SrTiO$_3$(100) substrates are used for the MBE growth of FeTe films. FeTe films grown on metallic SrTiO$_3$(100) are used in STM/S measurements, while those grown on insulating SrTiO$_3$(100) are used in ex situ electrical transport and scanning SQUID measurements.

Before MBE growth, all SrTiO$_3$(100) substrates are first soaked in hot deionized water ($\sim$80 °C) for 2 hours, then immersed in a $\sim$4.5\% HCl solution for 2 hours, and finally annealed at $\sim$974 $^\circ$C for 3 hours in a tube furnace with flowing oxygen. These treatments passivate and reconstruct the SrTiO$_3$(100) surface, making it suitable for the MBE growth of FeTe films. These heat-treated SrTiO$_3$(100) substrates are loaded into the MBE chambers and outgassed at $\sim$600 $^\circ$C for 1 hour before the MBE growth. High-purity Fe (99.995\%) and Te (99.9999\%) are co-evaporated from Knudsen effusion cells. The substrate temperature is kept at $\sim$330 $^\circ$C during the MBE growth. The growth rate is $\sim$0.2 unit cell (UC) per minute.

FeTe films with a nominal thickness of 40 UC ($\sim$25 nm) are employed in this study. The FeTe films in samples A-D are annealed under a Te flux at $\sim$280 $^\circ$C for 30 minutes (A), 20 minutes (B), 15 minutes (C), and 10 minutes (D), respectively, to remove excess Fe atoms and induce superconductivity. The FeTe film in sample E does not undergo any post-growth Te-annealing treatment and therefore remains non-superconducting. No capping layer is involved for the samples used in all measurements.

\subsection{Scanning SQUID measurements}
We mounted FeTe samples on a home-built thermal stage that was thermally anchored to the mixing chamber of our Bluefors dilution fridge, allowing temperature sweeps from 0.2 to 20 K. A Cernox thermometer was mounted underneath the sample, and a heater was positioned far from both the sample and thermometer to minimize thermal lag. The SQUID susceptometer was mounted on a separate stage controlled by the piezo positioners and scanners. The SQUID-sample distance $z$ during measurements was $\sim1 - 2\ \mathrm{\mu m}$.

The SQUID susceptometer used in this work consists of a pickup loop (inner/outer radii of 0.4/1 $\mathrm{\mu m}$) coupled to a SQUID circuit for detecting magnetic flux and a field coil (inner/outer radii of 1.5/3 $\mathrm{\mu m}$). To generate screening currents in FeTe, we applied an AC current with an amplitude of 0.2--0.5 mA at a frequency of about 200 Hz through the field coil using a lock-in amplifier, producing a local magnetic field of approximately 20--50 $\mathrm{\mu}$T. The SQUID approached the sample surface at a shallow angle of 3$^\circ$--5$^\circ$, making the applied local field predominantly out of the sample plane and therefore inducing in-plane screening currents. These screening currents coupled flux of opposite sign into the pickup loop, resulting in a reduction in the mutual inductance $\Delta M$ detected by the lock-in. For thin films with thickness $d$ much smaller than the London penetration depth $\lambda$, the change in mutual inductance $\Delta M$ can be modeled by \cite{kirtley2012scanning}
\begin{equation}\label{eq:mutual}
    \Delta M/M_\mathrm{FC}=-\frac{ad}{2\lambda^2(T)}\left( 1-\frac{2z}{\sqrt{a^2+4z^2}} \right),
\end{equation}
where $a$ is the effective field coil radius \cite{kirtley2016scanning}, and $z$ is the distance between the SQUID and the sample. Therefore, $\Delta M$ measured at a constant $z$ is proportional to $\lambda^{-2}$ and thus to the superfluid stiffness $K_s$. An optical image of the SQUID sensor and a schematic of the sample stage are provided in Supplementary Section S1.

\subsection{Electrical transport measurements}
The FeTe films grown on 3 mm$\times$10 mm heat-treated SrTiO$_3$(100) substrates are scratched into a Hall bar geometry with an effective area of $\sim$1 mm$\times$0.5 mm using a computer-controlled motorized probe station. The electrical contacts are made by pressing indium spheres on the Hall bar. The electrical transport measurements are conducted using a Physical Property Measurement System (PPMS, Quantum Design DynaCool, 1.7 K, 9 T). The excitation current is 1 $\mathrm{\mu A}$ in all temperature-dependent resistance measurements. To minimize oxidation, all samples are measured within 30 minutes after being taken out of the MBE chamber.

\subsection{STM/S measurements}
The STM/S measurements are performed in a Unisoku 1300 system with a base vacuum better than $2\times10^{-10}$ mbar. The system incorporates a single-shot 3He cryostat to achieve a base temperature of $\sim$310 mK. The maximum magnetic field is $\sim$11 T. Polycrystalline PtIr tips are used in our STM/S measurements. Before STM/S measurements on a 40-UC stoichiometric FeTe film, the PtIr tips are conditioned on an MBE-grown Ag film to ensure clean and stable tunneling characteristics. The d$I$/d$V$ spectra are obtained using the standard lock-in method by applying an additional small AC excitation voltage at a frequency $f$ = 987.5 Hz. All STM data are treated using standard functions in MATLAB, Python 3.9, and WSxM 5.0 software \cite{horcas2007wsxm}.

\subsection{Theoretical calculations}

To capture the tunneling spectrum observed in STM/S and the two-gap behavior, we theoretically model the stoichiometric FeTe following the tight-binding model in
Ref.~\cite{yan2026stoichiometric}.
First-principles density functional theory (DFT) calculations with experimentally extracted lattice constants show that bands near the Fermi level predominantly consist of Fe-$d$ orbitals. Thus, a single-particle tight-binding model projected to these $d$-orbitals can be obtained by Wannierizing the DFT bands to capture the low-energy physics.
The Wannier Hamiltonian on the $k_z=0$ plane is further unfolded to produce a 2D lattice with one Fe atom per unit cell \cite{eschrig2009tbmodel,wu2016topological}, yielding a single-particle tight-binding model on the so-called 1-Fe lattice.
In terms of the primitive lattice vectors of FeTe, $\mathbf{a}_\text{FeTe,1}$ and $\mathbf{a}_\text{FeTe,2}$, the primitive lattice vectors in the 1-Fe lattice are given by $\mathbf{a}_\text{Fe,1}=\frac{1}{2}\mathbf{a}_\text{FeTe,1}-\frac{1}{2}\mathbf{a}_\text{FeTe,2}$, $\mathbf{a}_\text{Fe,2}=\frac{1}{2}\mathbf{a}_\text{FeTe,1}+\frac{1}{2}\mathbf{a}_\text{FeTe,2}$.

On top of the single-particle bands, interaction terms are incorporated to capture the strong correlations in FeTe. The onsite repulsion is modeled by the Hubbard-Kanamori interaction \cite{gastiasoro2016unconventional,brydon2011magnetic,wu2016topological},
\begin{align}
H_{\mathrm{int,onsite}}
=&
\frac{U}{2}
\sum_{\mathbf R,\alpha,\sigma\ne\sigma^\prime}
 n_{\mathbf R\alpha\sigma}
 n_{\mathbf R\alpha\sigma^\prime}\nonumber\\
&+
\frac{2U'-J}{4}
\sum_{\mathbf R,\alpha\neq\beta,\sigma,\sigma'}
 n_{\mathbf R\alpha\sigma}
 n_{\mathbf R\beta\sigma'}
\nonumber\\
&
-4J
\sum_{\mathbf R,\alpha\neq\beta}
{\mathbf S}_{\mathbf R\alpha}
\cdot
{\mathbf S}_{\mathbf R\beta}
\nonumber\\
&+
\frac{J'}{2}
\sum_{\mathbf R,\alpha\neq\beta,\sigma}
c^\dagger_{\mathbf R\alpha\sigma}
c^\dagger_{\mathbf R\alpha\bar{\sigma}}
c_{\mathbf R\beta\bar{\sigma}}
c_{\mathbf R\beta\sigma},
\end{align}
where $c_{\mathbf{R}\alpha\sigma}^\dagger$ is the electron creation operator at lattice vector $\mathbf{R}$ of the 1-Fe lattice with spin $\sigma$ in orbital $\alpha$, $n_{\mathbf{R}\alpha\sigma} = c_{\mathbf{R}\alpha\sigma}^\dagger c_{\mathbf{R}\alpha\sigma}$ is the electron density operator, and $\mathbf{S}_{\mathbf{R}\alpha}=(\mathrm{S}_{\mathbf{R}\alpha,x},\mathrm{S}_{\mathbf{R}\alpha,y},\mathrm{S}_{\mathbf{R}\alpha,z})$ is the electron spin operator given by $\mathrm{S}_{\mathbf{R}\alpha,i}=\frac{1}{2}\sum_{\sigma,\sigma'}c^\dagger_{\mathbf{R}\alpha\sigma}c_{\mathbf{R}\alpha\sigma'}[s_i]_{\sigma,\sigma'}$ with $i=x,y,z$ and $s_i$ being the corresponding Pauli matrices.
The repulsive interaction contains the Hubbard interaction $U$, the Coulomb repulsive interaction $U'$ between electrons in different orbitals, the Hund's coupling $J$, and the pair-hopping energy $J'$.
A phenomenological NNNN AFM interaction is also added, which stabilizes the double-striped order~\cite{ma2009firstprinciples,Ducatman2014}, experimentally observed in FeTe samples with interstitial Fe \cite{Rodriguez2011,yan2026stoichiometric},
\begin{equation}
H_{\mathrm{int},J_3}
=
4J_3
\sum_{\mathbf R,\alpha,\boldsymbol{\delta}}
\mathbf{S}_{\mathbf R\alpha}
\cdot
\mathbf{S}_{\mathbf R+\boldsymbol{\delta},\alpha},
\end{equation}
where $\boldsymbol{\delta}\in\{\pm2\mathbf{a}_\text{Fe,1},\pm2\mathbf{a}_\text{Fe,2}\}$ is the displacement between the NNNN Fe atoms.
The interaction strengths for Fig.~\ref{fig:gap-structure} are set to
$U=262.5$~meV,
$J_3=30$~meV,
$U'=U-2J$, $J=\frac{U}{4}$, and $J'=J$.

Furthermore, attractive interactions between NN and NNN are added to induce the superconducting pairing~\cite{gastiasoro2016unconventional},
\begin{align}
H_{\mathrm{SC}}
=
-\sum_i V_i & \sum_{\mathbf R,\alpha,\boldsymbol{\delta}_i}
\frac{\Gamma_{\alpha}}{4} \left( c^\dagger_{\mathbf R\alpha\uparrow} c^\dagger_{\mathbf R+\boldsymbol{\delta}_i,\alpha\downarrow} - c^\dagger_{\mathbf R\alpha\downarrow} c^\dagger_{\mathbf R+\boldsymbol{\delta}_i,\alpha\uparrow} \right)
\nonumber\\
&\times
\left(
c_{\mathbf R+\boldsymbol{\delta}_i,\alpha\downarrow}
c_{\mathbf R\alpha\uparrow}
-
c_{\mathbf R+\boldsymbol{\delta}_i,\alpha\uparrow}
c_{\mathbf R\alpha\downarrow}
\right).
\end{align}
Here $i=2,3$ denote NN ($\boldsymbol{\delta}_2 \in \{\pm\mathbf{a}_\text{Fe,1},\pm\mathbf{a}_\text{Fe,2}\}$) and NNN ($\boldsymbol{\delta}_3 \in \{$ $\mathbf{a}_\text{Fe,1}+\mathbf{a}_\text{Fe,2}$, $\mathbf{a}_\text{Fe,1}-\mathbf{a}_\text{Fe,2}$, $-\mathbf{a}_\text{Fe,1}+\mathbf{a}_\text{Fe,2}$, $-\mathbf{a}_\text{Fe,1}-\mathbf{a}_\text{Fe,2}$ $\}$) pairings, respectively.
The data presented in Fig.~\ref{fig:gap-structure} use the parameters $V_2=200$~meV
and
$V_3=300$~meV.
The form factor is $\Gamma_\alpha=1$ for the $d_{xy}$, $d_{xz}$, and $d_{yz}$ orbitals, and $\Gamma_\alpha=0.5$ for $d_{z^2}$ and $d_{x^2-y^2}$~\cite{gastiasoro2016unconventional}.
To explore nodal $s$-wave pairing, we considered an NNNN attractive interaction term as well in the Supplementary Materials, with strength denoted by $V_4$ and pairing displacements of $\boldsymbol{\delta}_4\in\{\pm2\mathbf{a}_\text{Fe,1},\pm2\mathbf{a}_\text{Fe,2}\}$. See Supplementary Materials for details.

The total Hamiltonian includes all the single-particle terms, denoted by $H_0$, and interaction terms, plus a chemical potential $\mu$ that keeps the filling constant at six electrons per Fe,
\begin{equation}
H = H_0 - \sum_{\mathbf R,\alpha,\sigma} \mu n_{\mathbf R\alpha\sigma} +H_{\mathrm{int,onsite}} + H_{\mathrm{int},J_3} + H_{\mathrm{SC}}.
\end{equation}
The full Hamiltonian is solved self-consistently in momentum space under the mean-field approximation on a $24\times24$ lattice.
The solver adjusts $\mu$ so that the electron filling is always fixed to 6 electrons per Fe atom.
The mean-field approximation assumes translational invariance, as well as the absence of spin-flipping processes, and therefore $S_z$ is conserved.
Since $S_z$ is conserved and the superconducting pairing is spin singlet, the full BdG Hamiltonian decomposes into equivalent reduced blocks.
The resulting reduced BdG Hamiltonian $h_\text{BdG}(\mathbf{k})$ is defined in the reduced Nambu basis in momentum space $(c_\mathbf{k\alpha\uparrow},c^\dagger_\mathbf{-k\beta\downarrow})^T$, where $\alpha,\beta$ run over the five Fe $d$ orbitals.
In the Nambu basis, the particle-particle and hole-hole sectors of the mean-field Hamiltonian form the normal blocks, while the anomalous particle-hole and hole-particle blocks are responsible for superconducting pairing.
We refer to the Fermi surfaces of the normal block as the renormalized Fermi surfaces, which do not preserve the Luttinger count as the chemical potential is determined in the presence of pairing.
In addition to $S_z$ conservation, we enforce spin-$\mathrm{SU}(2)$ and spinful time-reversal symmetries in the solver.
The 2D low-energy interacting Hamiltonian $H$ also has point-group symmetries, including inversion and mirrors normal to the $\mathbf{a}_\text{FeTe,1}+\mathbf{a}_\text{FeTe,2}$, $\mathbf{a}_\text{FeTe,1}-\mathbf{a}_\text{FeTe,2}$ and $\mathbf{a}_\text{FeTe,1}$ directions. However, they are not enforced and may be spontaneously broken by the converged solutions. In particular, the nodal $d$-wave solutions presented spontaneously break the mirror symmetry normal to $\mathbf{a}_\text{FeTe,1}$ in the anomalous block, i.e., the particle-hole sector, of the mean-field Hamiltonian.
On the other hand, the nodal $s$-wave solution discussed in the Supplementary Material retains all the spatial symmetries.

Due to spin-$\mathrm{SU}(2)$ symmetry, spinful time-reversal symmetry implies spinless time-reversal symmetry of the reduced BdG Hamiltonian, represented by $\mathcal{T}=\mathcal{K}$ with $\mathcal{K}$ the complex conjugate, such that $h_\text{BdG}^*(\mathbf{k})=h_\text{BdG}(-\mathbf{k})$.
Similarly, the particle-hole symmetry of the full BdG Hamiltonian can also combine with spin-$\mathrm{SU}(2)$ to produce the particle-hole symmetry of the reduced BdG Hamiltonian, represented by $\tau_y\mathcal{K}$ where $\tau_y$ is the Pauli matrix defined on the particle-hole index of the reduced Nambu basis. The Hamiltonian satisfies $\tau_y h_\text{BdG}^*(\mathbf{k}) \tau_y=-h_\text{BdG}(-\mathbf{k})$.
The product of the spinless time-reversal and the spinless particle-hole symmetries further yields the spinless chiral symmetry of the reduced BdG Hamiltonian, defined by $\mathcal{C}=\tau_y$ such that $\mathcal{C}=\mathcal{C}^\dagger=\mathcal{C}^{-1}$.
It anticommutes with the reduced BdG Hamiltonian, i.e., $\mathcal{C} h_\text{BdG}(\mathbf{k})=-h_\text{BdG}(\mathbf{k})\mathcal{C}$.
Due to chiral symmetry, the nodal points can be unambiguously characterized by the chiral winding number~\cite{wen1989winding,volovik2007qpt,sato2011andreev,schnyder2011surface,chiu2016classification}.
The chiral winding number is defined around a closed loop in momentum space by
\begin{equation}
W[C]=\frac{i}{4\pi}\oint_C \operatorname{Tr} [\mathcal{C} h^{-1}_\text{BdG}(\mathbf{k}) \mathrm{d} h_\text{BdG}(\mathbf{k}) ].
\end{equation}
It counts the topological charge due to point nodes enclosed by the contour $C$.
Winding number calculations show that the nodes in Fig.~\ref{fig:gap-structure}(d) indeed host nodal charges of $\pm 1$ in the reduced Nambu space, establishing the existence of the nodal points protected by the chiral symmetry.

Across a pairing node with a chiral winding number of $\pm1$, the effective pairing must change sign on the interaction-renormalized Fermi surface.
Since the low energy effective model of FeTe contains multiple bands, the pairing function is defined by projecting onto the zero-energy states of the normal block of the reduced BdG Hamiltonian, as follows.
Let $u_\mathbf{k}$ be the zero-energy eigenvector (column vector) on the Fermi surface of the particle-particle block of $h_\text{BdG}(\mathbf{k})$, which has a dimension of five. Then, the spinless time-reversal symmetry allows us to derive projected pairing as $\Delta(\mathbf{k})=(u_\mathbf{k},\mathbf{0})^\dagger h_\text{BdG}(\mathbf{k}) (\mathbf{0},u_{-\mathbf{k}}^*) = (u_\mathbf{k},\mathbf{0})^\dagger h_\text{BdG}(\mathbf{k}) (\mathbf{0},u_{\mathbf{k}})$, where the $\mathbf{0}$ denotes the zero vector. This gives the effective pairing amplitude, as shown in Fig.~\ref{fig:gap-structure}(d).

Lastly, we discuss the calculation of the local DOS for comparison with the experimental data in Fig.~\ref{fig:gap-structure}. The theoretical local DOS in Fig.~\ref{fig:gap-structure}(e) is calculated by
$\frac{\mathrm{d}I}{\mathrm{d}V} \propto \sum_{n}\int \mathrm{d}^2\mathbf{k} \|u_{n\mathbf{k}}\|^2 \frac{\eta}{\pi}
[ \left(eV_{\text{bias}}-\varepsilon_{n\mathbf{k}}\right)^2+\eta^2
]^{-1}$,
where $\varepsilon_{n\mathbf{k}}$ is the $n$th eigenenergy at momentum $\mathbf{k}$, $u_{n\mathbf{k}}$ is the particle component of the eigenstate $(u_{n\mathbf{k}},v_{n\mathbf{k}})^T$, and $\eta=0.05$~meV is the broadening used.
To obtain a smooth curve without smearing out sharp features in the DOS, we take the converged mean-field Hamiltonian and solve for its spectrum on an extended hybrid momentum mesh.
The hybrid mesh consists of a coarse $240\times240$ momentum grid covering the first BZ and a fine grid around momenta for which the spectrum of the normal blocks of the converged MF BdG Hamiltonian lies within $\pm25$~meV of the Fermi level.
Here, the normal blocks are obtained by retaining all converged particle-number-conserving Hartree--Fock terms in $h_\text{BdG}$ while setting the anomalous Nambu off-diagonal blocks to zero.
The momentum points in this fine grid form a subset of the momentum grid of a $24000\times24000$ lattice.
In addition, we monitor the polarization towards superconducting pairing of the mean-field Hamiltonian,
defined at a bias voltage $V_\text{bias}$ by $P_\text{SC}=\sum_n \int \mathrm{d}^2\mathbf{k} \delta(\varepsilon_{n\mathbf{k}}-eV_\text{bias})  4|u^\dagger_{n\mathbf{k}} v_{n\mathbf{k}}|^2/\sum_n \int \mathrm{d}^2\mathbf{k} \delta(\varepsilon_{n\mathbf{k}}-eV_\text{bias})$ when $eV_\text{bias}$ is within the spectral support of the BdG Hamiltonian, and zero otherwise.
It shows that at the bias voltage of 10 mV, the pairing polarization is reduced to 0.12, compared to 0.63 at 5~mV. Hence, the effect of the superconducting interaction is mainly limited to below 10 mV.

\bibliographystyle{apsrev4-1-etal-title}
\bibliography{refs-JJAs.bib}

\clearpage
\onecolumngrid

\begin{center}
{\large\bfseries Supplementary Information:\\[0.25em]
Signatures of nodal superconductivity in stoichiometric FeTe\par}

\vspace{1.0em}

Cequn Li$^{1,*}$, Zi-Jie Yan$^{2}$, Yang Ge$^{3,4}$, Zihao Wang$^{2}$, Bing Xia$^{2}$, Stephen Paolini$^{2}$,\\
Pu Xiao$^{2}$, Lok-Kan Lai$^{2}$, Jiatao Song$^{2}$, Austin R. Kaczmarek$^{1}$, Lujin Min$^{5}$, Kenji Yasuda$^{5}$,\\
Peter J. Hirschfeld$^{3}$, Jiabin Yu$^{3,4}$, Cui-Zu Chang$^{2,\dagger}$, and Katja C. Nowack$^{1,6,\ddagger}$

\vspace{0.75em}

{\small
$^{1}$Laboratory of Atomic and Solid-State Physics, Cornell University, Ithaca, NY 14853, USA\\
$^{2}$Department of Physics, The Pennsylvania State University, University Park, PA 16802, USA\\
$^{3}$Department of Physics, University of Florida, Gainesville, FL 32611, USA\\
$^{4}$Quantum Theory Project, University of Florida, Gainesville, FL 32611, USA\\
$^{5}$Department of Applied and Engineering Physics, Cornell University, Ithaca, NY 14853, USA\\
$^{6}$Kavli Institute at Cornell for Nanoscale Science, Ithaca, NY 14853, USA\\[0.4em]
$^{*}$cl2775@cornell.edu \qquad
$^{\dagger}$cxc955@psu.edu \qquad
$^{\ddagger}$kcn34@cornell.edu\\[0.4em]
}
\end{center}

\vspace{0.5em}

\makeatletter
\newcommand{\sitableofcontents}{\section*{Contents}\@starttoc{sitoc}}
\newcommand{\sisection}[1]{\section{#1}\addcontentsline{sitoc}{section}{\protect\numberline{\thesection}#1}}
\newcommand{\sisubsection}[1]{\subsection{#1}\addcontentsline{sitoc}{subsection}{\protect\numberline{\thesubsection}#1}}
\makeatother

\setcounter{section}{0}
\setcounter{subsection}{0}
\setcounter{secnumdepth}{2}
\setcounter{equation}{0}
\setcounter{figure}{0}
\setcounter{table}{0}
\setcounter{page}{1}
\renewcommand{\thepage}{S\arabic{page}}
\renewcommand{\thesection}{S\arabic{section}}
\renewcommand{\theequation}{S\arabic{equation}}
\renewcommand{\thefigure}{S\arabic{figure}}

\makeatletter
\renewcommand*{\theHsection}{SI.\arabic{section}}
\renewcommand*{\theHsubsection}{SI.\arabic{section}.\arabic{subsection}}
\renewcommand*{\theHequation}{SI.\arabic{equation}}
\renewcommand*{\theHfigure}{SI.\arabic{figure}}
\renewcommand*{\theHtable}{SI.\arabic{table}}
\makeatother

\sitableofcontents

\clearpage

\sisection{Scanning SQUID setup}\label{sec:intro}

\begin{figure*}[tbph]
\centering \includegraphics[width=0.95\textwidth]{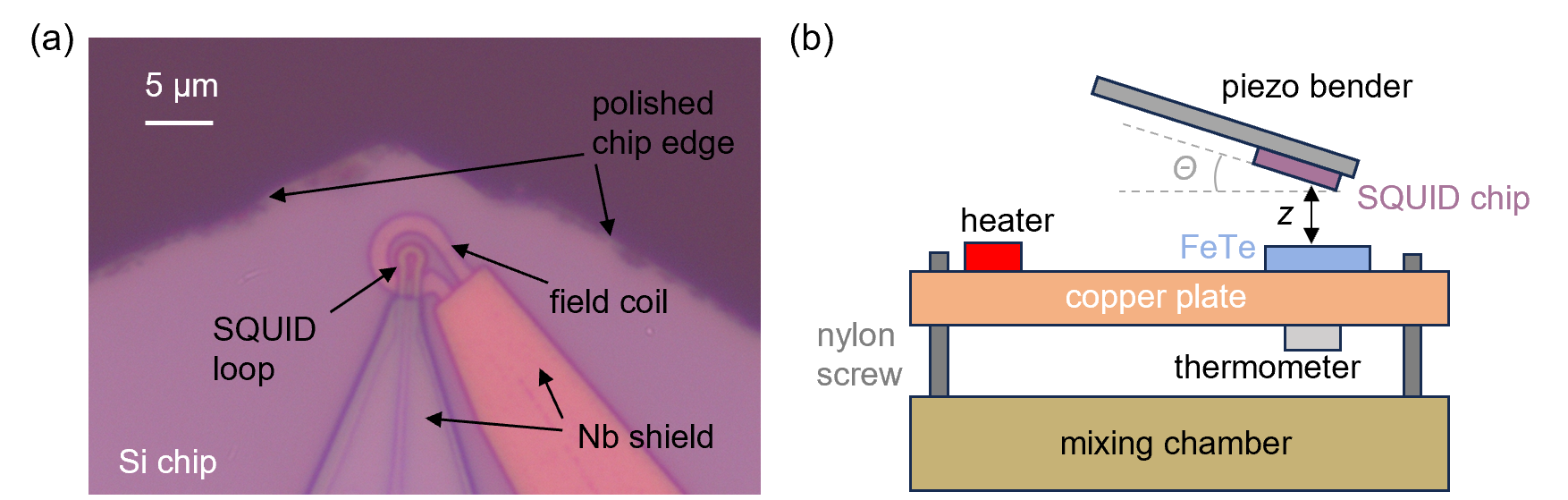}
\caption{\textbf{Experimental setup.} \textbf{a,} Optical image of the scanning SQUID susceptometer used in this work. \textbf{b,} Schematic of the setup for the penetration-depth measurement (not to scale). The SQUID chip sits on a piezo bender and approaches the sample at a small angle $\theta\approx5^\circ$.}
\label{fig:squid}
\end{figure*}

Figure \ref{fig:squid}(a) shows an optical image of the SQUID susceptometer used in this work. It consists of a pickup loop coupled to a SQUID circuit for detecting magnetic flux and a field coil that applies a local magnetic field by carrying an AC current \cite{huber2008gradiometric}. The pickup loop has inner and outer radii of 0.4 and 1 $\mathrm{\mu m}$, respectively, and the field coil has inner and outer radii of 1.5 and 3 $\mathrm{\mu m}$, respectively. Far from the sample, we measure the mutual inductance $M_\mathrm{FC}=246\pm1\ \mathrm{\phi_0/A}$ between the field coil and the pickup loop by detecting the SQUID flux response at the drive frequency using a lock-in amplifier. As the SQUID approaches the superconducting sample, screening currents are induced over a length scale set by the field coil geometry. These screening currents couple flux of opposite sign into the pickup loop, resulting in a reduction in the mutual inductance $\Delta M$. For thin films with thickness $d$ much smaller than the London penetration depth $\lambda$, $\Delta M$ is proportional to the superfluid stiffness $K_s$ of the superconductor \cite{kirtley2012scanning}. Details of the modeling are provided in Section~\ref{sec:fitting_lambda}.

Figure \ref{fig:squid}(b) schematically illustrates the experimental setup for our measurements. The FeTe sample is mounted on a copper plate thermally anchored to the mixing chamber (MXC) of our dilution fridge. To sweep the temperature above $T_c\sim12$ K of FeTe without heating the MXC, we use nylon screws as the thermal bridges between the copper plate and MXC. The base temperature of the sample plate is approximately 200 mK. A Cernox thermometer is mounted underneath the sample, and a heater sits far away from both the thermometer and the sample to minimize thermal lag. The SQUID susceptometer is mounted on a separate stage whose motion is controlled by the piezo positioners and scanners. The SQUID-sample separation $z$ during scanning is typically $\sim1\ \mathrm{\mu m}$.

\clearpage
\sisection{Images of vortices and background field compensation}\label{sec:vortex}

\begin{figure*}[tbph]
\centering \includegraphics[width=0.95\textwidth]{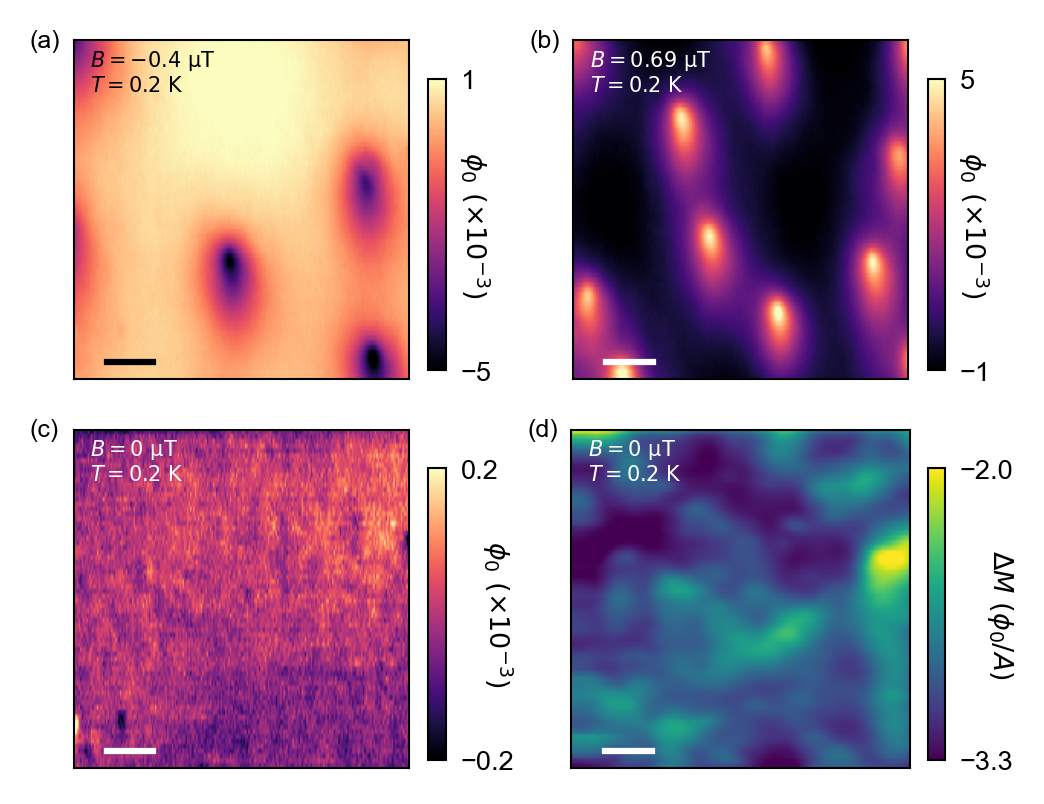}
\caption{\textbf{Imaging vortices in sample A.} \textbf{a}--\textbf{b,} Images of vortices under a finite background magnetic field, detected by the dc flux channel of our SQUID. \textbf{c,} DC flux image at nearly zero magnetic field in the same field of view. \textbf{d}, Diamagnetic response map acquired in the same field of view.}
\label{fig:vortex}
\end{figure*}

After cooling the sample through $T_c$ under a finite background magnetic field, vortices appear within our field of view (Figs. \ref{fig:vortex}(a) and (b)), detected by the dc-flux channel of our SQUID. By adjusting the background field using a Helmholtz coil inside the fridge, we reduced the vortex density to nearly zero (Fig. \ref{fig:vortex}(c)). This ensures that no nearby vortices move in response to the AC field‑coil excitation and therefore do not contribute spurious signals to $\Delta M$.

We note the presence of weak magnetic textures even at zero magnetic field in the dc-flux images in sample A (Fig. \ref{fig:vortex}(c)). Their stray‑field amplitudes are roughly 30 times smaller than the stray-field amplitude of an isolated superconducting vortex, and they persist up to $T=25$ K, well above the superconducting $T_c$. Moreover, they show no clear spatial correlation with the diamagnetic‑response image (Fig. \ref{fig:vortex}(d)), indicating that their origin is unrelated to superconductivity.  The origin of these magnetic textures remains unknown. In particular, our measurements do not determine where within the film the magnetic signals originate or the microscopic source of the magnetism. Further studies are needed to clarify their origin.

\clearpage
\sisection{Temperature evolution of the diamagnetic-response map in sample A}\label{sec:temp-series}

\begin{figure*}[tbph]
\centering \includegraphics[width=0.8\textwidth]{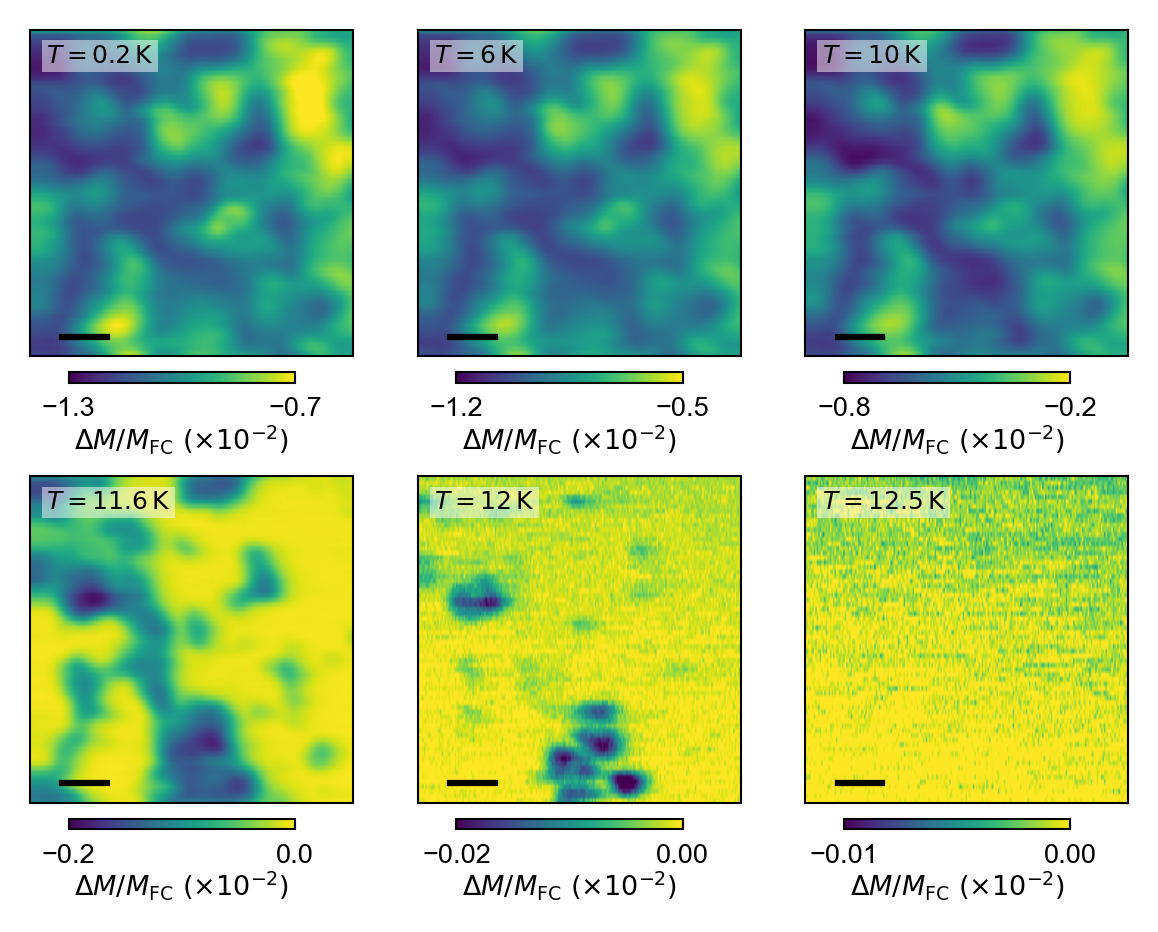}
\caption{\textbf{Temperature evolution of diamagnetic response.} An image series of the diamagnetic response in sample A acquired from 0.2 K to 12.5 K. Darker regions correspond to more negative $\Delta M$ and therefore stronger superfluid stiffness. These regions turn normal at slightly higher temperatures. The $T_c$ map as shown in the main text is extracted from these images and images taken at additional temperatures.  $T_c$ is defined as the temperature corresponding to a small threshold $\Delta M/M_{\mathrm{FC}}=-5\times10^{-5}$ at each pixel. Scale bar, 20 $\mathrm{\mu m}$.}
\label{fig:temp series}
\end{figure*}

\clearpage
\sisection{Spatial correlations and distribution of \texorpdfstring{$T_c$}{Tc}}\label{sec:correlation}

\begin{figure*}[tbph]
\centering \includegraphics[width=1\textwidth]{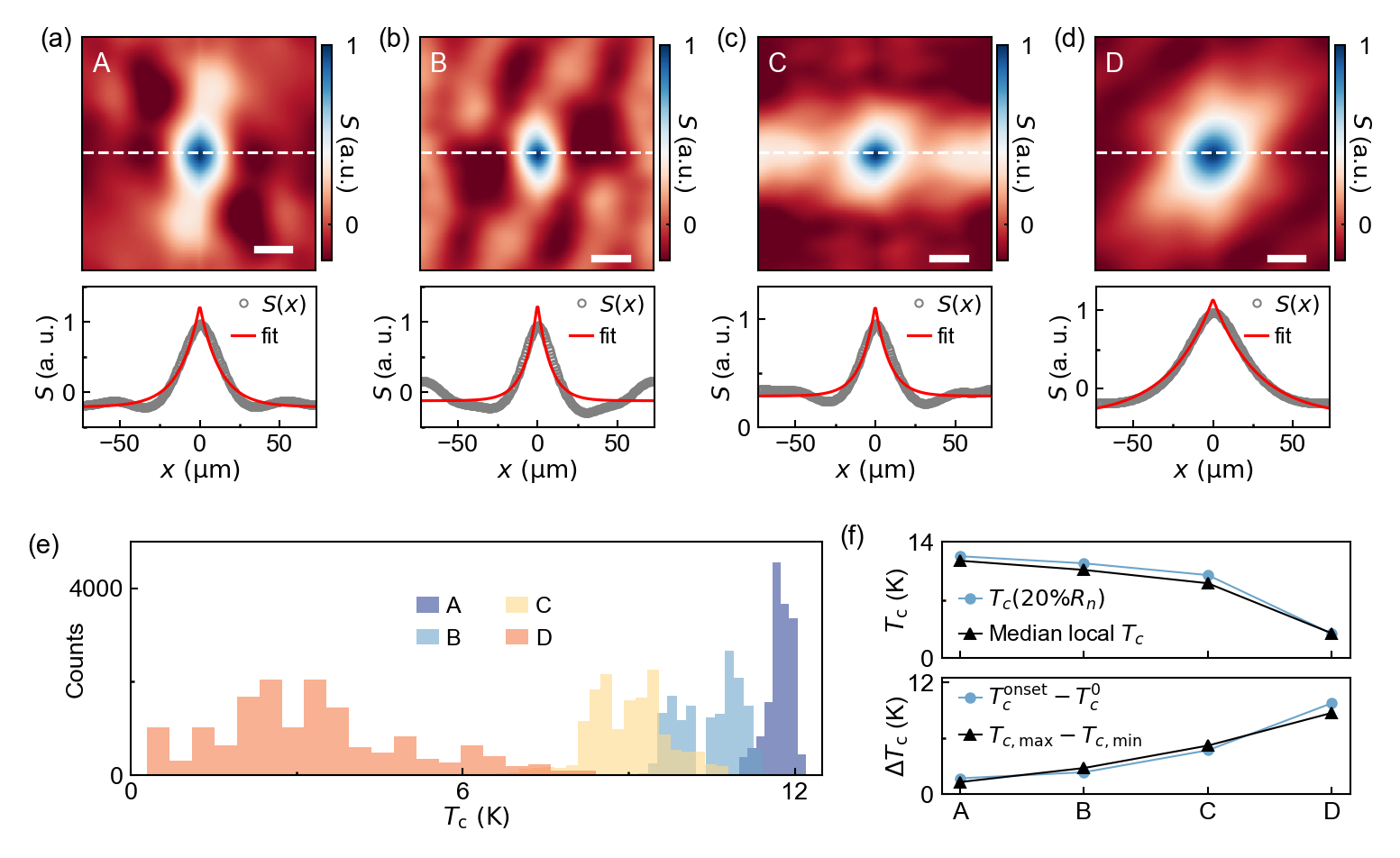}
\caption{\textbf{Autocorrelations and distribution of $T_c$.} \textbf{a--d,} Autocorrelations of the diamagnetic-response maps shown in Fig. 2(a) and Figs. 4(a)–(c) of the main text. Scale bar, 20 $\mathrm{\mu m}$. The bottom panels show line cuts of the autocorrelation (gray dots) together with fits to $S(x)=S_0e^{-x/\zeta}$ (red lines). Values of $\zeta$ extracted from the fits are 12.7, 8.1, 9, and 27.8 $\mathrm{\mu m}$ for samples A--D, respectively. \textbf{e,} Histogram of the local $T_c$ values for samples A--D, extracted from the $T_c$ maps in Fig. 2(b) and Figs. 4(d)–(f) of the main text. \textbf{f,} Top: comparison between the transport‑defined $T_c$ at 20\% of the normal-state resistance $R_n$(15 K) (blue dots) and the median local $T_c$ extracted from the histograms (black triangles). Bottom: comparison between the width of the superconducting transition in transport, $T_c^{\mathrm{onset}}-T_c^0$, and the peak-to-peak variation of the local $T_c$ distribution, $T_{c,\mathrm{max}}-T_{c,\mathrm{min}}$.}
\label{fig:autocorrelation}
\end{figure*}

Figures \ref{fig:autocorrelation}(a)--(d) show the autocorrelations of the diamagnetic-response maps in Fig. 2(a) and Figs. 4(a)--(c) of the main text. The central peaks demonstrate that neighboring points in the spatial distribution of the diamagnetic response are correlated. The autocorrelation can be fit to an exponential decay, $S(x)=S_0e^{-x/\zeta}$, where $\zeta$ gives the characteristic domain size. Across samples A–D, we obtain $\zeta\approx8$--30 $\mathrm{\mu m}$, larger than the dimensions of our SQUID sensor, whose field coil has inner and outer radii of 1.5 and 3~$\mathrm{\mu m}$, respectively, and whose pickup loop has inner and outer radii of 0.4 and 1~$\mathrm{\mu m}$, respectively. The absence of secondary peaks in the autocorrelations indicates that the domains lack long‑range periodic order.

Figure \ref{fig:autocorrelation}(e) plots the histogram of the local $T_c$ values, which systematically broadens and shifts to lower $T_c$ from samples A to D, consistent with the trend observed in the transport measurements in Fig. 1(b) of the main text. Figure \ref{fig:autocorrelation}(f) demonstrates close agreement between the transport‑defined $T_c$ at 20\% of the normal-state resistance $R$(15 K) (blue dots) and the median local $T_c$ extracted from the histograms (black triangles). Moreover, the width of the superconducting transition in transport, $T_c^{\mathrm{onset}}-T_c^0$, matches the peak-to-peak variation of the local $T_c$ distribution. These results establish a direct connection between the macroscopic transport properties and the underlying spatial inhomogeneity of FeTe.

\clearpage
\sisection{Surface morphology of stoichiometric FeTe thin films}\label{sec:surface-morphology}

\begin{figure*}[tbph]
\centering \includegraphics[width=1\textwidth]{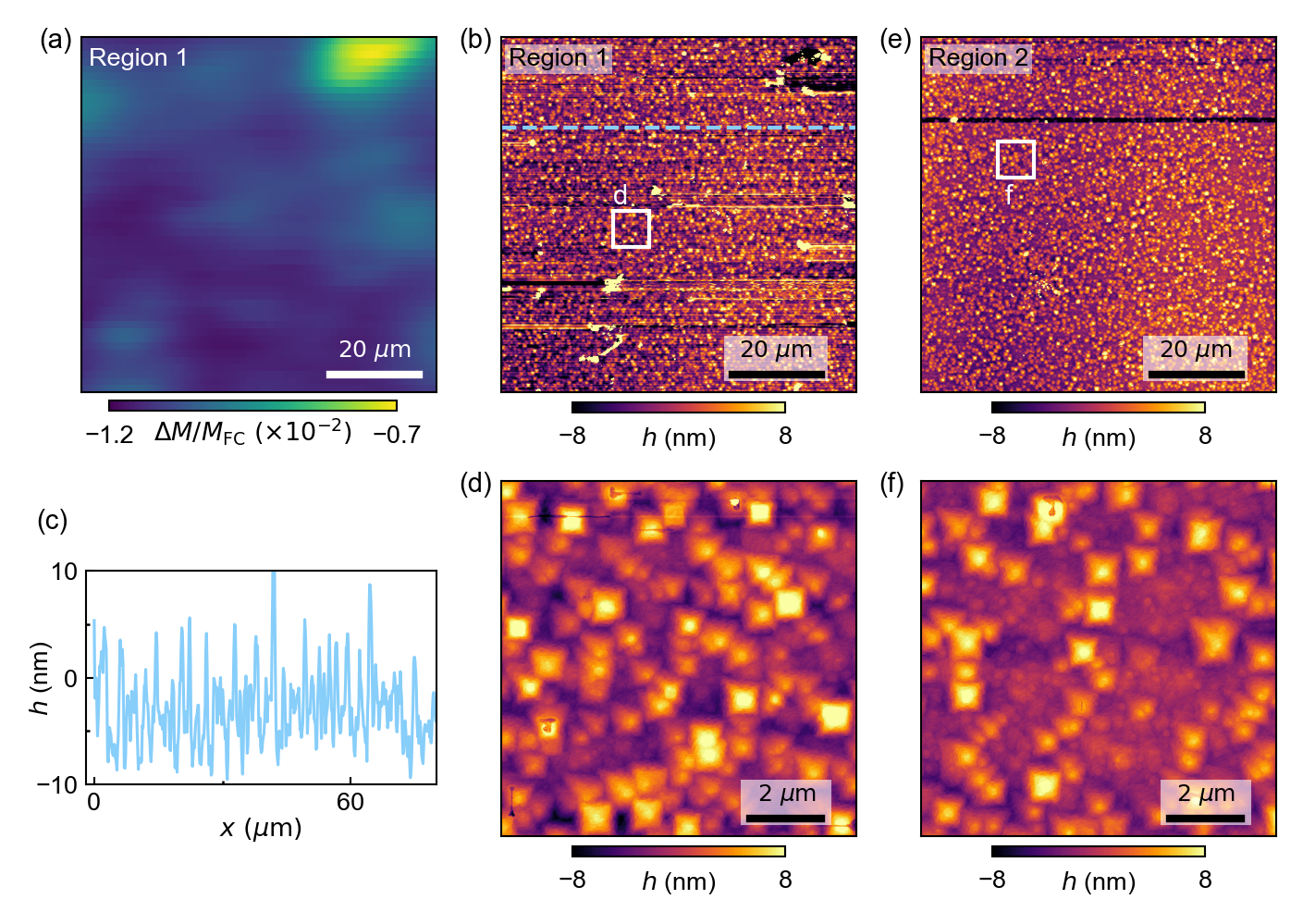}
\caption{\textbf{Comparison between the diamagnetic-response map and surface morphology.} \textbf{a,} Diamagnetic-response map taken at $T=0.2$ K in sample A. \textbf{b,} Atomic-force-microscopy image acquired from approximately the same region at room temperature. \textbf{c,} Height profile along the blue dashed line in \textbf{b}, showing a peak-to-peak surface roughness of about 15 nm. \textbf{d,} Zoomed-in atomic-force-microscopy image corresponding to the white box in \textbf{b}. \textbf{e,} Atomic-force-microscopy image taken from another region at room temperature, and \textbf{f} is the corresponding zoomed-in view.}
\label{fig:afm}
\end{figure*}

Figures \ref{fig:afm}(a) and (b) compare the diamagnetic-response map with the atomic force microscopy image taken from approximately the same region of sample A. The two images show little correlation, aside from the damaged area in the upper‑right corner. The superfluid stiffness varies on a length scale of about 10 $\mathrm{\mu m}$, whereas the characteristic length scale of the surface morphology is only $\sim0.6~\mathrm{\mu m}$ (see the zoomed-in image in Fig. \ref{fig:afm}(d)). Similar topography is observed in another region that was not measured by SQUID (see Figs. \ref{fig:afm}(e) and (f)). None of the topography images show features on a length scale comparable to those observed in Fig. 2(a) of the main text or in Fig. \ref{fig:chi}.

\clearpage
\sisection{Dependence of \texorpdfstring{$\Delta M(T)$}{Delta M(T)} on SQUID-sample distance}\label{sec:height-dependence}

\begin{figure*}[tbph]
\centering \includegraphics[width=0.8\textwidth]{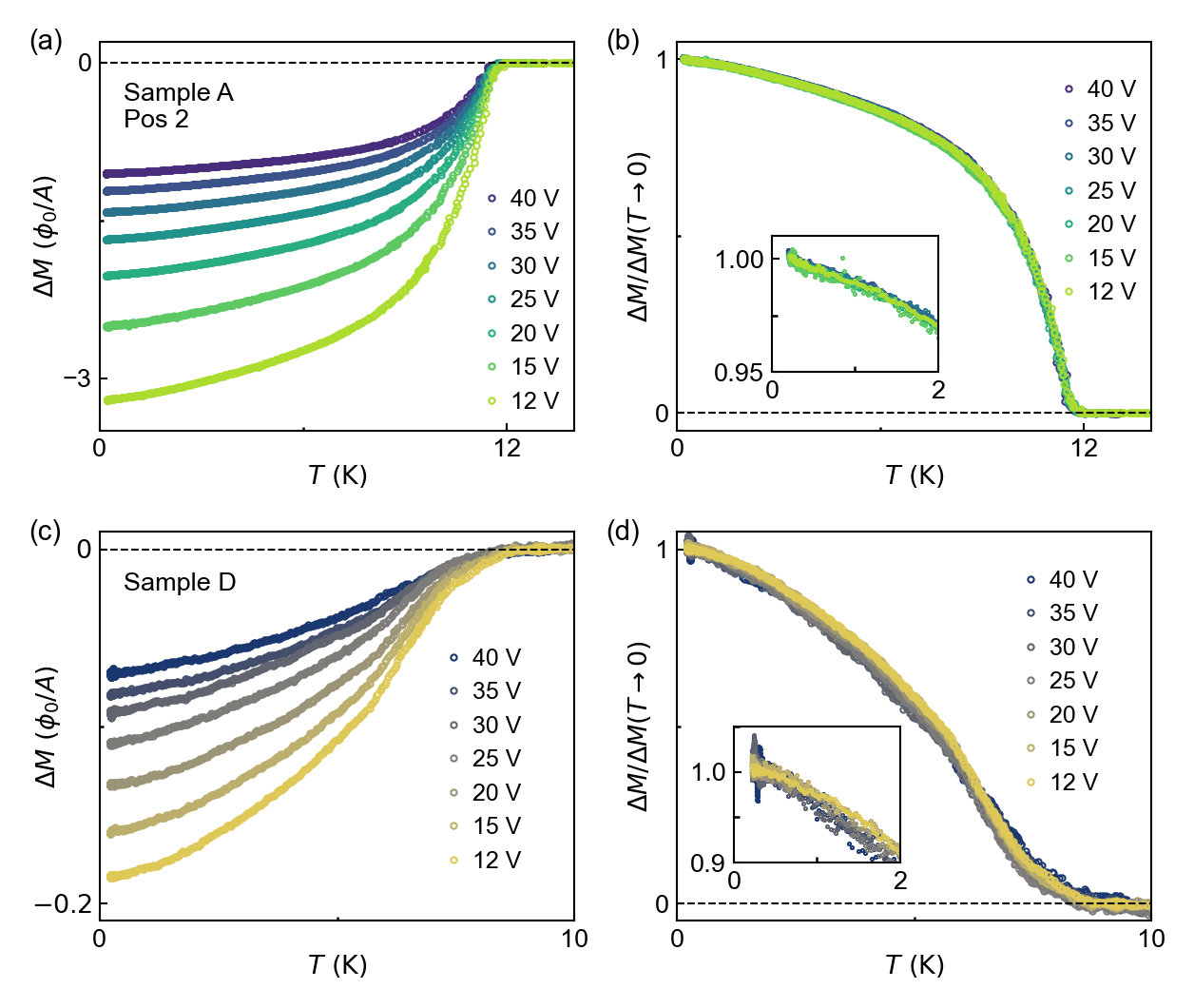}
\caption{\textbf{a,} $\Delta M(T)$ measured with different dc voltages $V_z-V_{td}$ applied to the piezo bender. The SQUID-sample distance is $z\approx 0.06\ \mathrm{\mu m/V}\times( V_z-V_{td})$. \textbf{b,} Normalized $\Delta M(T)/\Delta M(T=0)$ corresponding to the data in \textbf{a}. \textbf{c,} $z$-dependent $\Delta M(T)$ and \textbf{d,} $\Delta M(T)/\Delta M(T=0)$ in sample D.}
\label{fig:z dependence}
\end{figure*}

In the Pearl limit $\lambda\gg d$ with $d$ the film thickness, we have a simple relation between the London penetration depth $\lambda$ and the mutual inductance change $\Delta M$, i.e.,
\begin{equation}\label{eq:lambda}
    \frac{\lambda(T)}{\lambda(T=0)} =\sqrt{\frac{\Delta M(T=0)}{\Delta M(T)}},
\end{equation}
which is independent of field-coil radius $a$ and SQUID-sample distance $z$ \cite{kirtley2012scanning}. Figure \ref{fig:z dependence} shows $\Delta M(T)$ measured with the same SQUID at different $z$. The normalized $\Delta M(T)/\Delta M(0)$ curves collapse onto one another at multiple positions across all samples, confirming that the FeTe thin films operate in the Pearl limit.

\clearpage
\sisection{Examination of the relevant experimental parameters}\label{sec:Characterization of the experimental parameters}

\sisubsection{Dependence of \texorpdfstring{$\Delta M(T)$}{Delta M(T)} on field-coil excitation amplitude and frequency}\label{sec:FC-dependence}

\begin{figure*}[tbph]
\centering \includegraphics[width=0.8\textwidth]{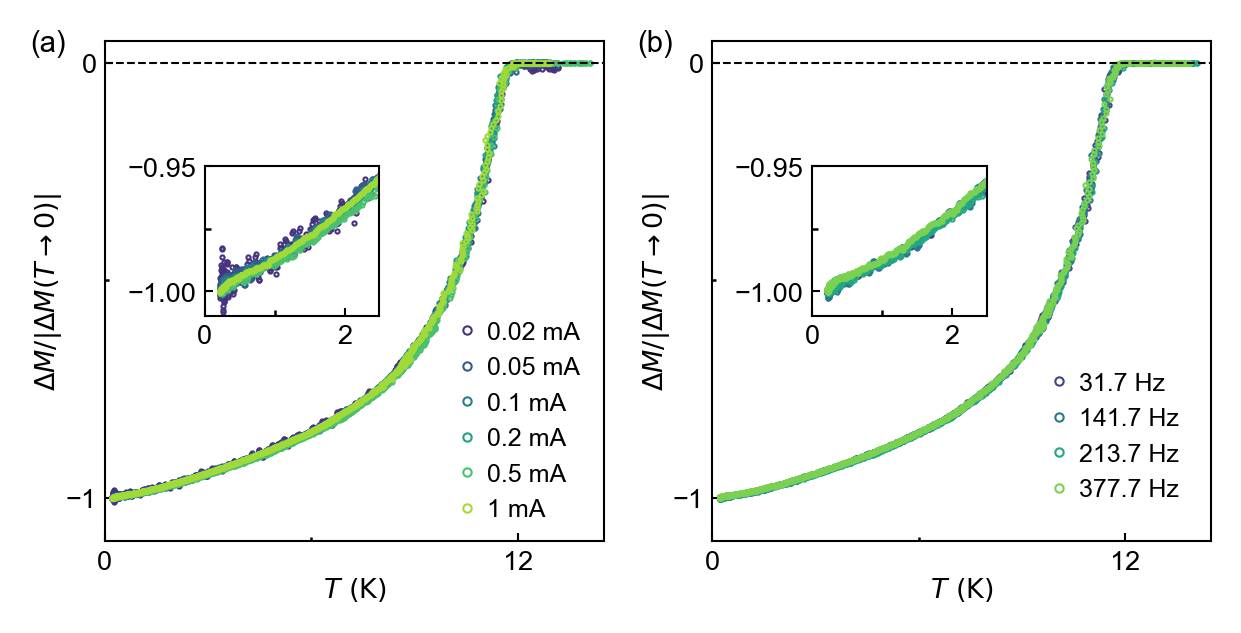}
\caption{\textbf{a,} $\Delta M(T)$ measured using different field-coil excitation amplitudes with a frequency of 213.7 Hz. Data were obtained at location 2 in sample A. \textbf{b,} $\Delta M(T)$ measured using different field-coil excitation frequencies with an amplitude of 0.5 mA.}
\label{fig:FC excitation}
\end{figure*}

We varied both the excitation amplitude and frequency, and Fig. \ref{fig:FC excitation} shows that the resulting $\Delta M(T)$ curves collapse onto one another. This confirms that the diamagnetic signal remains in the linear‑response regime over the range of excitation amplitudes used here. For the data presented in the main text, we used excitation amplitudes of 0.2–0.5 mA at a frequency of approximately 200 Hz.

\begin{figure*}[tbph]
\centering \includegraphics[width=0.5\textwidth]{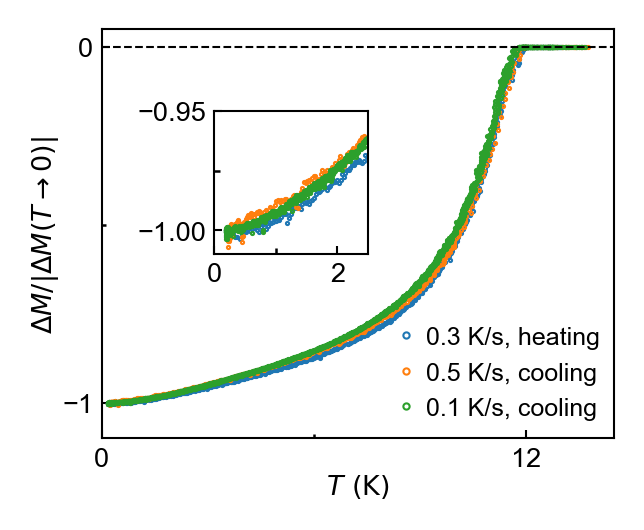}
\caption{\textbf{Dependence of $\Delta M(T)$ on the temperature-sweep direction and rate.}}
\label{fig:thermalization}
\end{figure*}

\clearpage
\sisubsection{Sample thermalization}\label{sec:thermalization}
A non-saturating behavior at low temperature can arise from thermal lag. We have examined this possibility by sweeping the temperature in opposite directions and at various rates. As shown in Fig. \ref{fig:thermalization}, the resulting $\Delta M(T)$ curves agree with one another, showing that thermal lag is negligible near the base temperature.

\begin{figure*}[tbph]
\centering \includegraphics[width=0.8\textwidth]{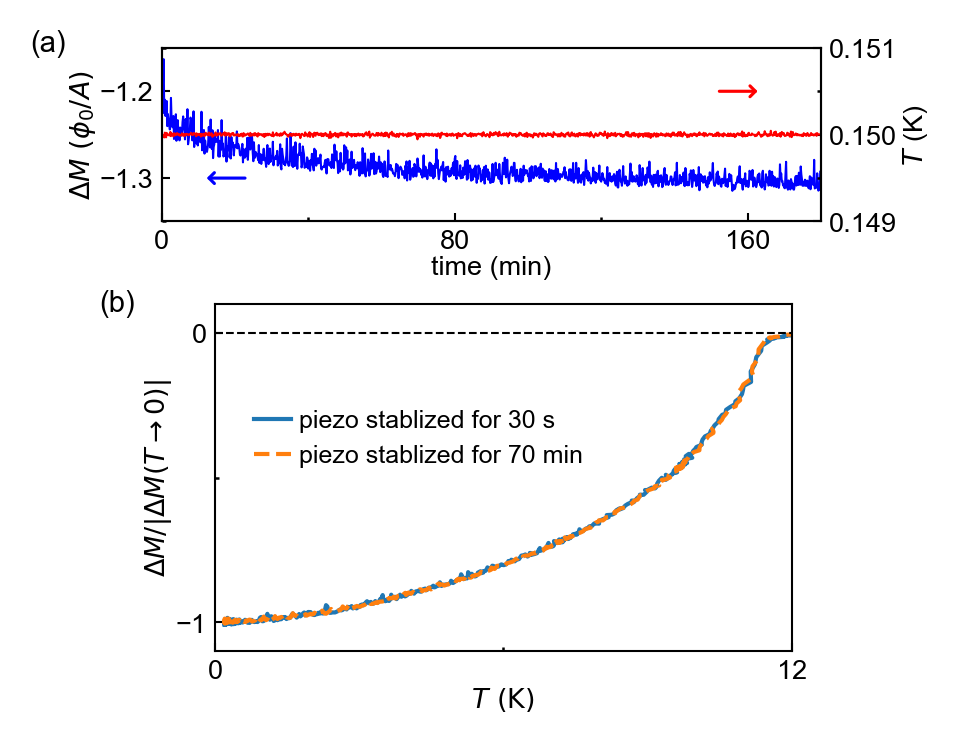}
\caption{\textbf{Effect of piezo drift.} \textbf{a,} Time trace of $\Delta M$ and $T$ after the SQUID is parked near the sample surface. A slight drift occurs during the first $\sim$ 15 min, after which the piezo scanner stabilizes. The sample temperature is 150 mK throughout the duration of the measurement. \textbf{b,} Comparison of $\Delta M(T)$ measured after short and long piezo-stabilization times.}
\label{fig:pz drift}
\end{figure*}

\begin{figure*}[tbph]
\centering \includegraphics[width=0.97\textwidth]{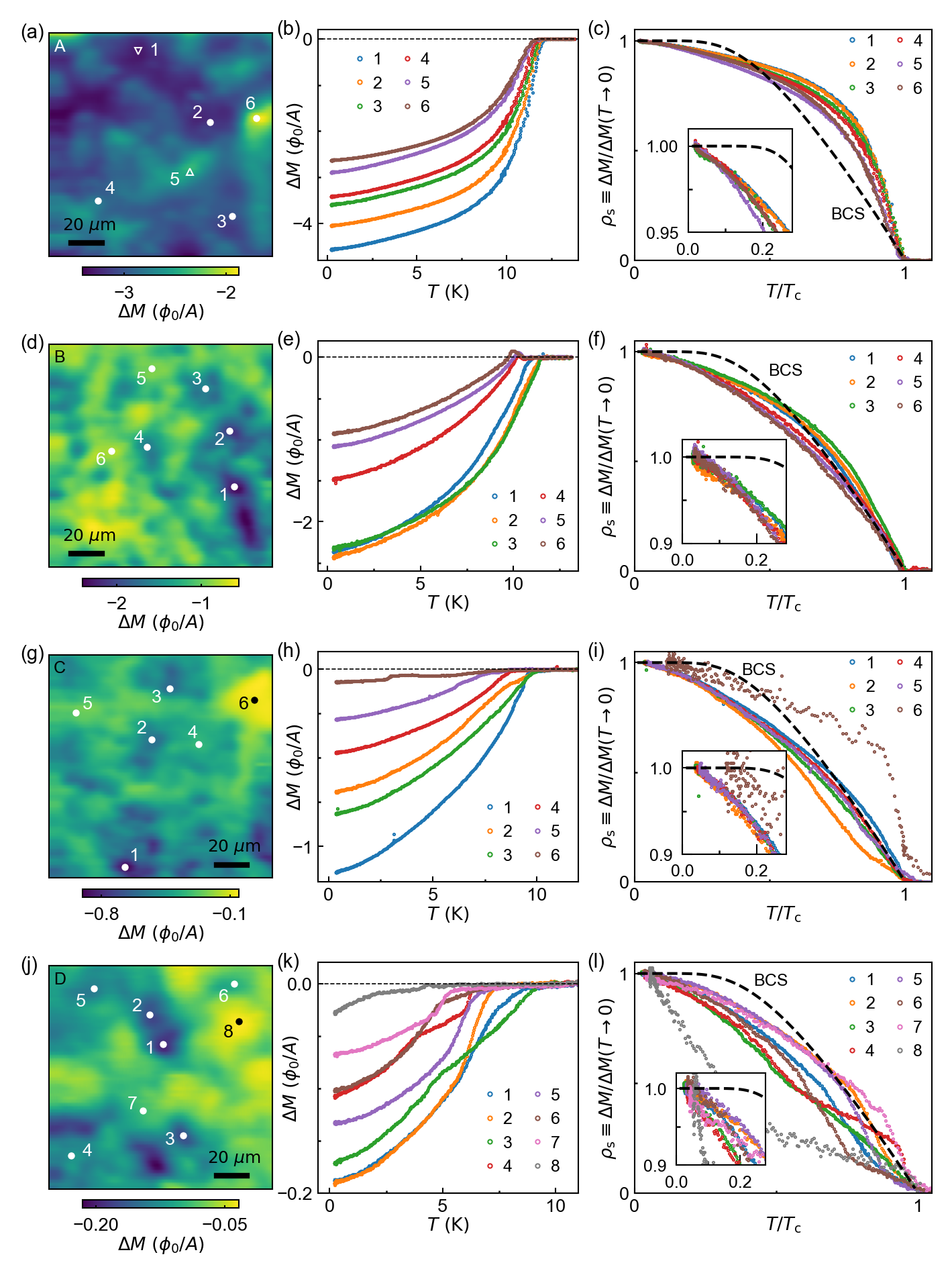}
\caption{\textbf{a,} Diamagnetic-response map at $T=0.2$ K, \textbf{b,} $\Delta M(T)$ at selected positions, and \textbf{c,} normalized superfluid stiffness $\rho_s$ versus $T/T_c$ for sample A. \textbf{d}--\textbf{f,} Same set of measurements for sample B. \textbf{g}--\textbf{i,} Data obtained from sample C. \textbf{j}--\textbf{l,} Data from sample D.}
\label{fig:chi}
\end{figure*}

\clearpage
\sisubsection{Piezo drift characterization}\label{sec:Piezo}
Our SQUID is mounted on a piezo scanner, and its position is controlled by applying a dc voltage to the scanner. We characterized the scanner's drift along the z direction by parking the SQUID $\sim$1 $\mathrm{\mu m}$ above a superconducting FeTe sample. As shown in Fig. \ref{fig:pz drift}(a), the magnitude of $\Delta M$ continues to change during the first $\sim$15 min after the piezo voltage is set, indicating that the scanner is still slowly approaching the sample. After $\sim30$ minutes, $\Delta M$ saturates. We estimate the maximum drift along the z-direction to be about 200 nm. To minimize the impact of piezo drift on the low-temperature data, we typically sweep temperature starting from above $T_c$, with each full sweep taking about 50 min. In Fig. \ref{fig:pz drift}(b), we compare $\Delta M(T)$ measured after short and long piezo-stabilization times and find no discernible difference between the two.

\sisection{Additional data in samples A--D}\label{sec:additionaldata}

Figure \ref{fig:chi} presents additional $\Delta M(T)$ curves measured at various locations across samples A--D. Most regions in samples A--C exhibit $T_c\geq7$ K, and their normalized superfluid stiffness $\rho_s(T)$ behaves consistently at low temperatures. In contrast, the behavior of $\rho_s(T)$ varies substantially, with signs of double transitions across sample D, suggesting that our SQUID averages signals from multiple regions.

\clearpage
\sisection{Power-law fits to \texorpdfstring{$\Delta\lambda(T)$}{lambda(T)} over different temperature ranges}\label{sec:power-law}

\begin{figure*}[tbph]
\centering \includegraphics[width=1\textwidth]{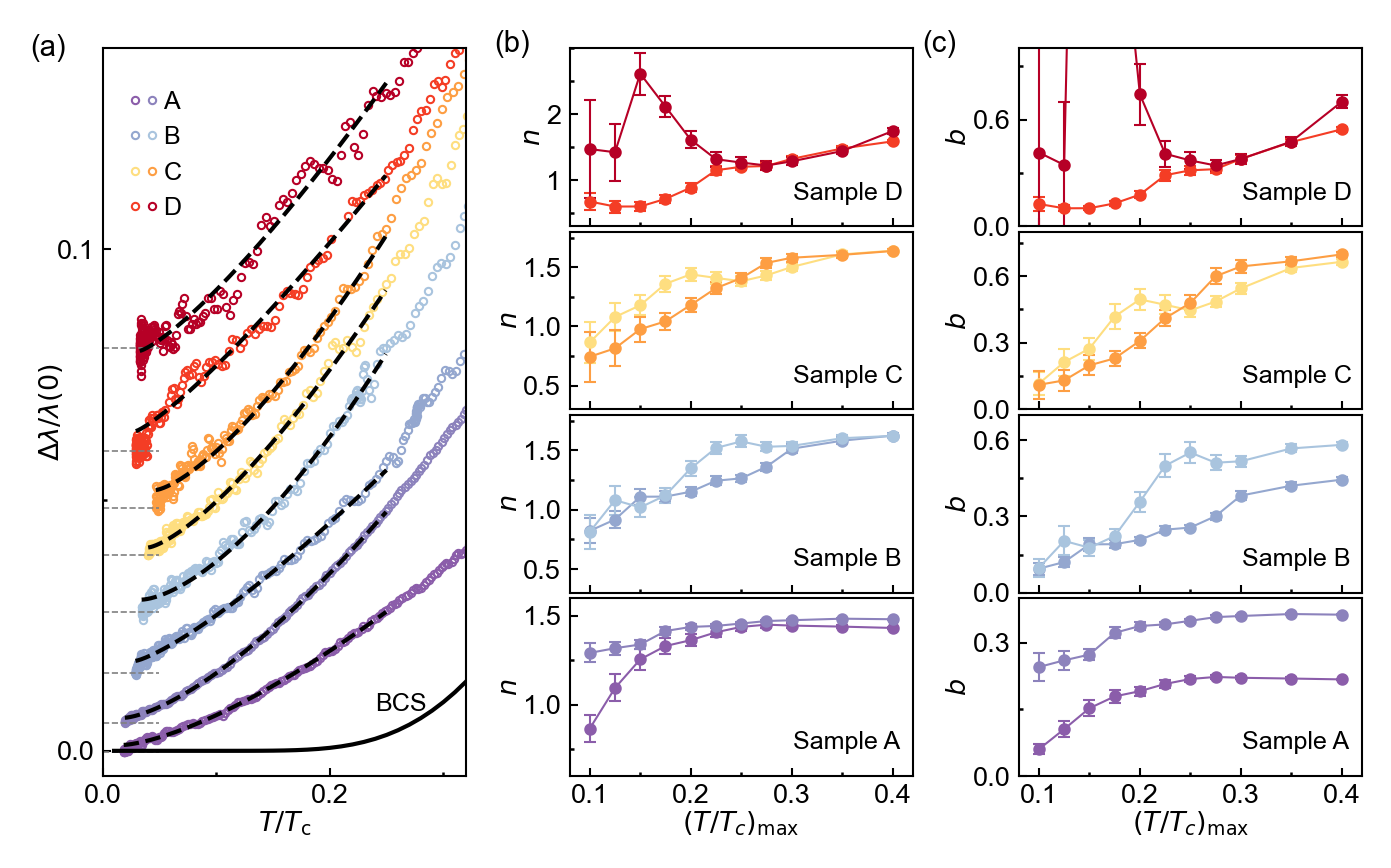}
\caption{\textbf{Power-law fits to the penetration depth in FeTe.} \textbf{a,} Same $\Delta\lambda/\lambda(0)$ data as in Fig. 3(a) of the main text. The black dashed lines show fits to $\Delta\lambda/\lambda(0)=b(T/T_c)^n$ for $T<0.25\ T_c$. The extracted exponent $n$ depends on the temperature range used for the fit. \textbf{b,} Fitted $n$ as a function of $(T/T_c)_\mathrm{max}$, obtained by varying the upper temperature limit $(T/T_c)_\mathrm{max}$ of the fitting window. \textbf{c,} Fitted pre-factor $b$ as a function of $(T/T_c)_\mathrm{max}$.}
\label{fig:power-law-fit}
\end{figure*}

 We perform power-law fits to $\Delta\lambda(T)$ over different temperature ranges and present the extracted exponent $n$ in Fig.~\ref{fig:power-law-fit}. We focus on sample A, where $n$ approaches $n\approx1$ near zero temperature and gradually saturates at $\sim1.5$ as the fitted temperature range increases. As a consistency check, we estimate the mean free path $l$ using the Drude expression $l=\hbar k_F/(n_{2D}e^2R_s)$. Taking the Fermi momentum $k_F\approx0.25\ \mathring{\mathrm{A}}^{-1}$ \cite{Lin2026,xu2026reversible}, two-dimensional (2D) normal-state carrier density $n_{2D}\approx10^{16}\ \mathrm{cm}^{-2}$ obtained from Hall measurements at $T=30$~K, and sheet resistance $R_s\approx100\ \Omega$ \cite{yan2026stoichiometric}, we find $l$ to be on the order of 1 nm, comparable to the superconducting coherence length $\xi\approx2$ nm \cite{xu2026reversible,yan2026stoichiometric}, placing stoichiometric FeTe between the clean and dirty limits.

\begin{figure*}[tbph]
\centering \includegraphics[width=0.6\textwidth]{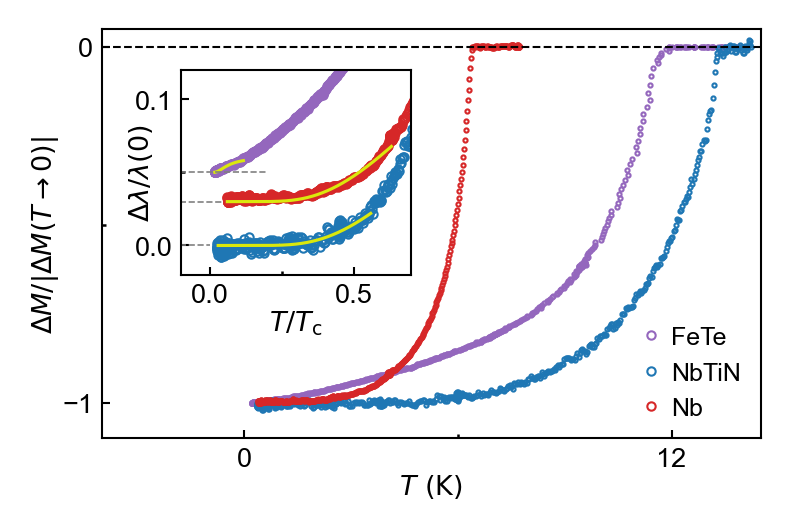}
\caption{\textbf{Exponential fits to penetration depth in FeTe, NbTiN, and Nb.} Normalized $\Delta M(T)/|\Delta M(T=0)|$ obtained for FeTe, NbTiN, and Nb. The inset shows the change in the normalized London penetration depth $\Delta\lambda/\lambda(0)$ versus $T/T_c$ and the corresponding exponential fits (yellow solid lines). Gray dashed lines indicate the vertical offsets applied to each curve. The extracted superconducting gaps are $\Delta_0=0.11\pm0.01$ meV, $2.55\pm0.15$ meV, and $1.18\pm0.03$ meV for FeTe, NbTiN, and Nb, respectively.}
\label{fig:exponential fit}
\end{figure*}

\clearpage
\sisection{Exponential fits to FeTe, Nb, and NbTiN data}\label{sec:expo}
For a fully gapped superconductor,  the change in the London penetration depth should exhibit an exponential temperature dependence \cite{prozorov2006magnetic}, i.e.,
\begin{equation}\label{eq:exp}
    \Delta\lambda(T)\propto T^{-1/2}e^{-\frac{\Delta_0}{k_BT}},
\end{equation}
where $\Delta_0$ is the zero-temperature superconducting gap. In Fig. \ref{fig:exponential fit}, we fit $\Delta\lambda/\lambda(0)$ versus $T/T_c$ measured in 25-nm-thick NbTiN and 20-nm-thick Nb films for $T<0.6T_c$ to Eq. \ref{eq:exp} and obtained $\Delta_0=2.55\pm0.15$ meV, and $1.18\pm0.03$ meV, respectively. These are consistent with 2.26 meV and 1.2 meV reported for NbTiN and Nb, respectively, in the literature \cite{hong2013terahertz,lemberger2007penetration}.

In addition, we fit the FeTe data below 0.1 $T_c$ to Eq. \ref{eq:exp} to estimate an upper bound for $\Delta_{min}$. The fit is poor because the data deviate from the simple exponential dependence. Nonetheless, we obtain $\Delta_0=0.11\pm0.01$ meV.

\clearpage
\sisection{Modeling of \texorpdfstring{$\rho(T)$}{rho(T)} for different gap models}\label{sec:fitting_rho}

\begin{figure*}[tbph]
\centering \includegraphics[width=1\textwidth]{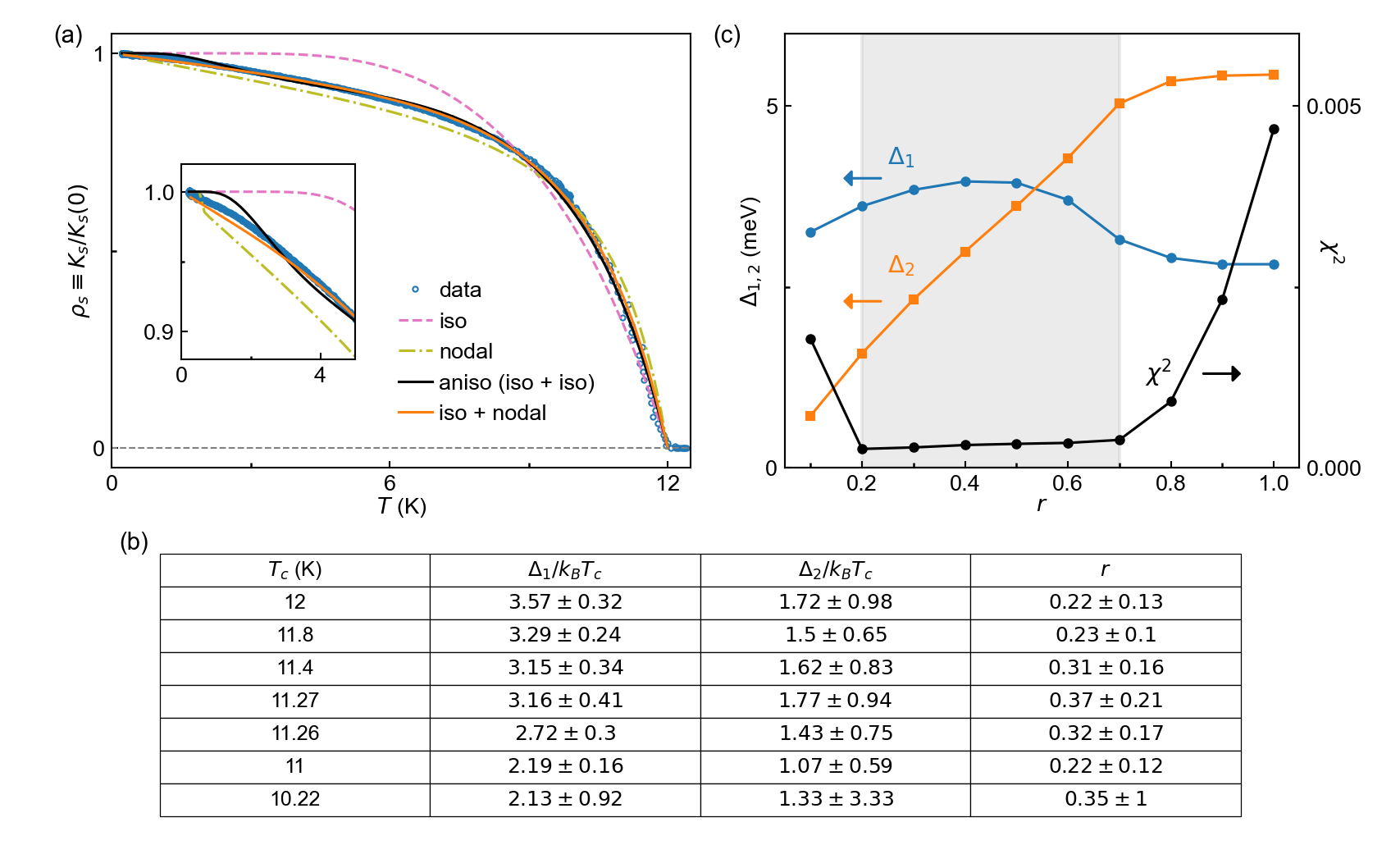}
\caption{\textbf{Fitting of the gaps.} \textbf{a,} Comparison between experimental data and fits. The isotropic (pink) and nodal (green) fits yield $\Delta_0=2.76\pm0.11$ meV and $\Delta_0=5.29\pm0.3$ meV, respectively. The anisotropic fit (black),  approximated as a weighted combination of two isotropic gaps using Eq. \ref{eq:two-gap}, gives $\Delta_1=3.33\pm0.08$ meV, $\Delta_2=0.61\pm0.09$ meV, and $r=0.29\pm0.02$. \textbf{b,} Parameters extracted from the best iso+nodal fits to the data for different $T_c$, as shown in Fig. 4(g) of the main text. Results obtained from the global fits are indicated in blue. \textbf{c,} Best-fit gap parameters \(\Delta_1\) and \(\Delta_2\) and the corresponding residual sum of squares \(\chi^2\) as functions of the weighting factor \(r\). For each fixed \(r\), \(\Delta_1\) and \(\Delta_2\) are optimized by nonlinear least-squares fitting. The gray shaded region denotes the range of parameters that provides comparably good fits to the data.
}
\label{fig:gap fit}
\end{figure*}

We model the temperature-dependent normalized superfluid stiffness $\rho_s$ using \cite{prozorov2006magnetic}
\begin{equation}\label{eq:rho_s}
    \rho_s(T)=\left(\frac{\lambda_0}{\lambda(T)}\right)^2=1-\frac{1}{2\pi k_BT}\int^{2\pi}_0\int^\infty_0\cosh^{-2}\left(\frac{\sqrt{\epsilon^2+\Delta^2(T,\varphi)}}{2k_BT}\right)d\epsilon d\varphi,
\end{equation}
where $\varphi$ is the in-plane azimuthal angle assuming a cylindrical Fermi
surface and $\Delta(T,\varphi)=\Delta(T)g(\varphi)$ is the gap. The temperature dependence of the gap amplitude is modeled within the $\alpha$-model framework \cite{johnston2013elaboration}, in which
\begin{equation}\label{eq:gap}
\Delta(T)=\alpha k_BT_c\frac{\Delta_{\mathrm{BCS}}(T)}{\Delta_{\mathrm{BCS}}(0)},
\end{equation}
where $\alpha$ is an adjustable parameter and $\Delta_{\mathrm{BCS}}(T)$ is the weak-coupling BCS gap obtained by numerically solving \cite{tinkham2004introduction}
\begin{equation}\label{eq:BCS}
\ln\left(T_c/T\right)=\int_0^\infty\left[\frac{\tanh\left(x\right)}{x}-\frac{\tanh\left(\sqrt{x^2+(\Delta_{\mathrm{BCS}}/2k_BT)^2}\right)}{\sqrt{x^2+(\Delta_{\mathrm{BCS}}/2k_BT)^2}}\right]dx.
\end{equation}
This approach allows $\Delta(T)$ to retain the BCS form while accommodating deviations of $\Delta(0)/k_B T_c$ from the weak-coupling value of 1.764 \cite{johnston2013elaboration}. The angular dependence $g(\varphi)$ characterizes the gap symmetry. For an isotropic \textit{s}-wave superconductor, $g(\varphi)=1$. A nodal gap with two-fold symmetry can be described by $g(\varphi)=\sin(2\varphi)$ \cite{prozorov2006magnetic}, whereas an anisotropic yet nodeless gap can be approximated as a weighted combination of two isotropic gaps. Figure \ref{fig:gap fit}(a) presents single-gap fits obtained using different $g(\varphi)$. None of them capture the experimental data at the low-temperature end.

Then we use a two-band $\alpha$-model adopted from the literature \cite{li2016superfluid,johnston2013elaboration}:
\begin{equation}\label{eq:two-gap}
    \rho_s=(1-r)\rho_{s1}(\Delta_1)+r\rho_{s2}(\Delta_2),
\end{equation}
where $0\leq r\leq1$ specifies the relative weight of the two bands, which contribute superfluid stiffnesses $\rho_{s1}$ and $\rho_{s2}$ with gaps $\Delta_{1}$ and $\Delta_{2}$, respectively. We treat $\Delta_1$ to be isotropic and $\Delta_2(T,\varphi)=\Delta_2(T)\sin(2\varphi)$. We assume a common $T_c$ for both gaps because no secondary transition or kink is observed in the data. The resulting fit agrees with the data well. Figure \ref{fig:gap fit}(b) summarizes the fit parameters obtained from the best fits to the data shown in Fig. 4(g) of the main text.

However, the fit parameters are not independently well constrained. To illustrate this, we fixed \(r\) at a series of values and optimized \(\Delta_1\) and \(\Delta_2\) at each fixed \(r\) using nonlinear least-squares fitting. The fit quality was quantified by the residual sum of squares, \(\chi^2=\sum_i[\rho_s^{\mathrm{exp}}(T_i)-\rho_s^{\mathrm{fit}}(T_i)]^2\). The resulting best-fit \(\Delta_1\), \(\Delta_2\), and \(\chi^2\) are plotted as functions of \(r\) in Fig.~\ref{fig:gap fit}(c). The gray shaded region denotes the parameter range that yields comparably good fits to the data, corresponding to $\Delta_1=$3.15–3.96~meV and $\Delta_2=$1.58–5.03~meV.

\clearpage
\sisection{Fitting of \texorpdfstring{$\lambda$}{lambda}}\label{sec:fitting_lambda}

We estimate the London penetration depth $\lambda$ using
\begin{equation}\label{eq:mutual_vs_Lambda}
    \Delta M/M_\mathrm{FC}=-\frac{ad}{2\lambda^2(T)}\left( 1-\frac{2z}{\sqrt{a^2+4z^2}} \right)=-\frac{L_{geo}d}{2\lambda^2(T)},
\end{equation}
where $M_\mathrm{FC}=246\pm1\ \mathrm{\phi_0/A}$ is the mutual inductance between SQUID and field coil measured far away from the sample, $a$ is the effective field coil radius (typically taken as the ``Ketchen" radius $a=\sqrt{(a_{in}^2+a_{out}^2)/2}$ \cite{kirtley2016scanning}), $z$ is the SQUID-sample distance, and $L_{geo}=a\left( 1-\frac{2z}{\sqrt{a^2+4z^2}} \right)$ is the geometric coefficient \cite{kirtley2012scanning}. The height $z$ depends on the piezo-bender voltage $V-V_{td}$, i.e., $z=\beta(V-V_{td})+z_0$, where $\beta$ is the bender constant, and $V_{td}$ and $z_0$ are the piezo-bender voltage and height when the SQUID chip first makes mechanical contact with the sample. The uncertainty in the fitting of $\lambda(0)$ is dominated by the error associated with $\beta$ and $z_0$.

\begin{figure*}[tbph]
\centering \includegraphics[width=1\textwidth]{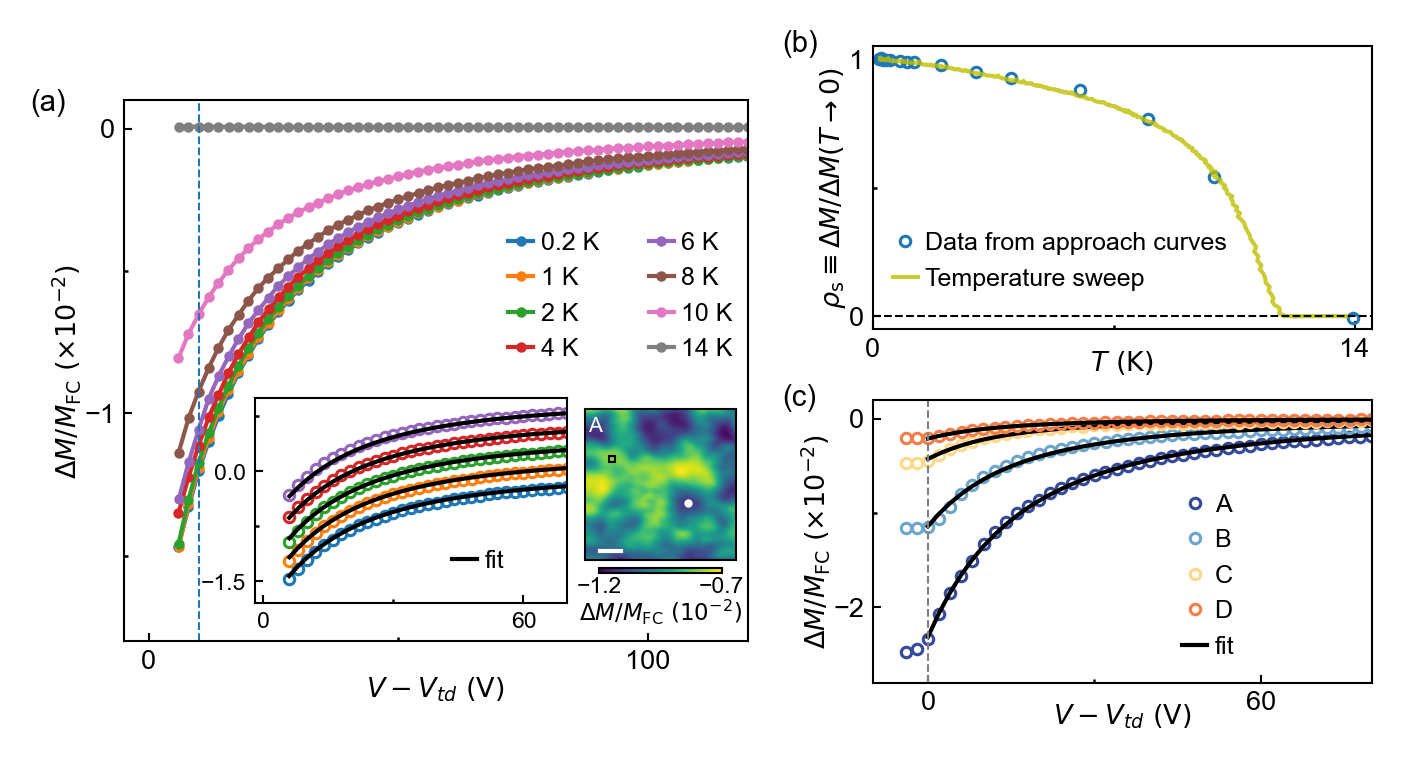}
\caption{\textbf{Fitting the penetration depth from the approach curves.} \textbf{a,} $\Delta M/M_\mathrm{FC}$ versus piezo-bender voltage $V-V_{td}$ obtained at different temperatures in sample A. The SQUID chip contacts the sample surface at $V-V_{td}=0$. The left inset shows fits to the data from 0.2 K to 6 K. Curves are vertically offset for clarity. Data were acquired at the location marked by the white dot in the diamagnetic response map (right inset). \textbf{b,} Comparison between the normalized $\Delta M/\Delta M(0)$ extracted from the line cut (blue dashed line) in \textbf{a} and the $\Delta M(T)/\Delta M(0)$ curve obtained from a temperature sweep. Data are normalized to their respective 0.2 K values. We calculated $\lambda(T)/\lambda(0)$ using Eq. \ref{eq:lambda}, and its behavior below 6 K is well described by $\lambda(T)=\lambda(0)[1+0.006\times T^{1.44}]$. \textbf{c,} Representative approach curves from samples A--D and the corresponding fits.}
\label{fig:approach}
\end{figure*}

Figure \ref{fig:approach}(a) presents $\Delta M/M_{\mathrm{FC}}$ as the SQUID approaches the sample at different temperatures in sample A. These measurements produce results consistent with the $\Delta M(T)$ curves obtained from a continuous temperature sweep at a constant $z$ (Fig. \ref{fig:approach}(b)). We fit the approach curves from 0.2 K to 6 K simultaneously using Eq. \ref{eq:mutual_vs_Lambda}, imposing the constraint $\lambda(T)=\lambda(0)[1+b\ T^n]$, where $b=0.006$ and $n=1.44$ are determined using the data in Fig. \ref{fig:approach}(b). Table \ref{tab:para} summarizes the fitting results, and we extract $\lambda(0)=512\pm27$ nm.

Because the same SQUID and piezo bender were used for samples A--D, $a$ and $\beta$ are identical across all measurements. Fitting the approach curves in Fig. \ref{fig:approach}(c) by imposing the ``Ketchen'' radius $a=2.37\ \mathrm{\mu m}$ and $\beta=63\pm3$ nm/V yields $L_{geo}(z=z_0)$ of $0.51\pm0.01$, $0.61\pm0.02$, $0.86\pm0.1$, and $0.81\pm0.04\ \mathrm{\mu m}$ for samples A--D. The modest sample‑to‑sample variations in $L_{geo}$ arise from small differences in $z_0$ due to sample‑dependent SQUID alignment. In Fig. 4(h) of the main text, we compute $K_s(0)=\frac{\upsilon|\Delta M(T\to0,z=z_0)|}{M_\mathrm{FC}}/L_{geo}(z=z_0)$ with  $\upsilon=\hbar^2/(2\mu_0k_Be^2)\approx1.24~\mathrm{K\cdot cm}$, where $\Delta M(T\to0,z=z_0)$ is measured from the approach curves for each data point, with error bars reflecting the uncertainties in $L_{geo}(z=z_0)$.

The temperature sweeps of $\Delta M(T)$ were taken at $z$($V-V_{td}=12$ V). To obtain $K_s(T)$ as plotted in Fig. 4(g) of the main text, we scale $\frac{|\Delta M(T)|}{M_\mathrm{FC}}/L_{geo}(z=z_0)$ by the factor $1.24~\mathrm{K\cdot cm}\times\frac{\Delta M(V=V_{td})}{\Delta M(V=V_{td}+12\mathrm{V})}$, determined from the data in Fig. \ref{fig:approach}(c) for each sample. We apply the same procedure to convert the diamagnetic response maps in Fig. 2(a) and Figs. 4(a)--(c) in the main text into $K_s(0)$, shown as the gray dots in Fig. 4(h) of the main text.

\begin{table}[t]
\centering
\begin{tabular}{p{2.5cm} p{2.5cm} p{2.5cm} p{2.5cm}}
\hline\hline
\ & $\lambda(0)\ (\mathrm{nm})$ & $\beta\ (\mathrm{nm/V})$ & $z_0\ (\mathrm{\mu m})$ \\ \hline
$a=2.37\ \mathrm{\mu m}$ & $512\pm27$ & $63\pm3$ & $1.76\pm0.13$ \\
\hline\hline
\end{tabular}
\caption{\textbf{Parameters obtained by fitting the approach curves in Fig. \ref{fig:approach}(a) to Eq. \ref{eq:mutual_vs_Lambda}.} The radius $a$ is taken to be the effective ``Ketchen" radius, $a=\sqrt{(a_{in}^2+a_{out}^2)/2}=2.37\ \mathrm{\mu m}$, where $a_{in}$ and $a_{out}$ are the inner and outer radii of the field coil, respectively~\cite{kirtley2016scanning}.}
\label{tab:para}
\end{table}

\clearpage
\sisection{Comparing data with predictions from a phase-fluctuation model}\label{sec:PF}

For comparison with a phase-fluctuation (PF) scenario, we consider the phenomenological model from Ref. \cite{khvalyuk2024near}, which expresses the phase-fluctuation (PF)-induced temperature dependence of $\lambda$ as
\begin{equation}\label{eq:PFmodel}
    \frac{\Delta\lambda(T)}{\lambda(0)}=\left(\frac{T_c}{T_0}\right)^n\left(\frac{T}{T_c}\right)^n,
\end{equation}
where the temperature scale $T_0\propto K_s(0)$. In  amorphous InO$_\mathrm{x}$ thin films, $T_0\approx5K_s(0)$ \cite{khvalyuk2024near}. For $n=1$, the ratio $T_c/T_0$ determines the slope of the resulting linear dependence.

\begin{figure*}[tbph]
\centering \includegraphics[width=0.5\textwidth]{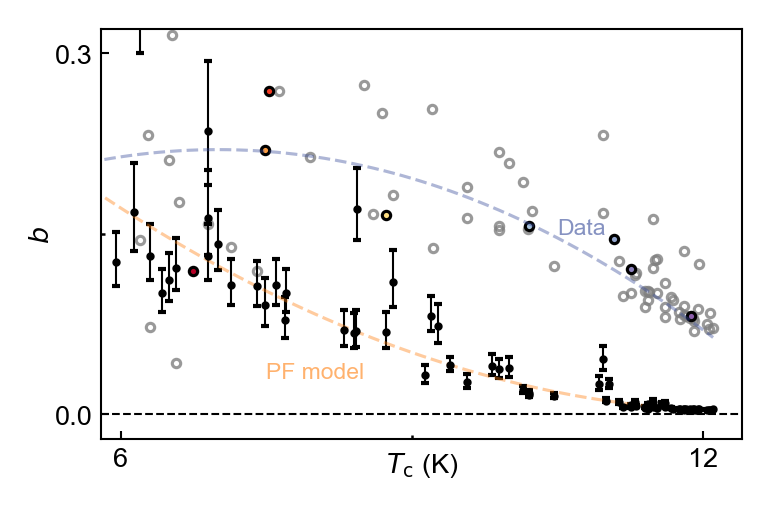}
\caption{Comparison between the extracted slope $b$ (color-coded and gray dots) from the linear fits in Fig. 3(a) of the main text and the calculation using $b=T_c/5K_s(0)$ (black dots) as a function of the local $T_c$. Orange and blue dashed lines are guides to the eye.}
\label{fig:PFmodel}
\end{figure*}

We compare the slope $b$ extracted from the data in Fig. 3(a) (gray and color-coded dots) with the expectation from the PF model assuming $T_0\approx5K_s(0)$ (black dots), as shown in Fig. \ref{fig:PFmodel}. Because $K_s(0)\gg T_c$ near $T_c=12$ K, PF-induced corrections are strongly suppressed and the calculated slope is much smaller than our observation. In contrast, as $T_c/K_s(0)$ increases rapidly with decreasing $T_c$, the calculated slope becomes comparable to the extracted $b$, indicating that disorder and inhomogeneity play an increasingly important role.

\clearpage
\sisection{Large-scale tunneling spectrum in stoichiometric FeTe}\label{sec:STM}

\begin{figure*}[tbph]
\centering \includegraphics[width=0.5\textwidth]{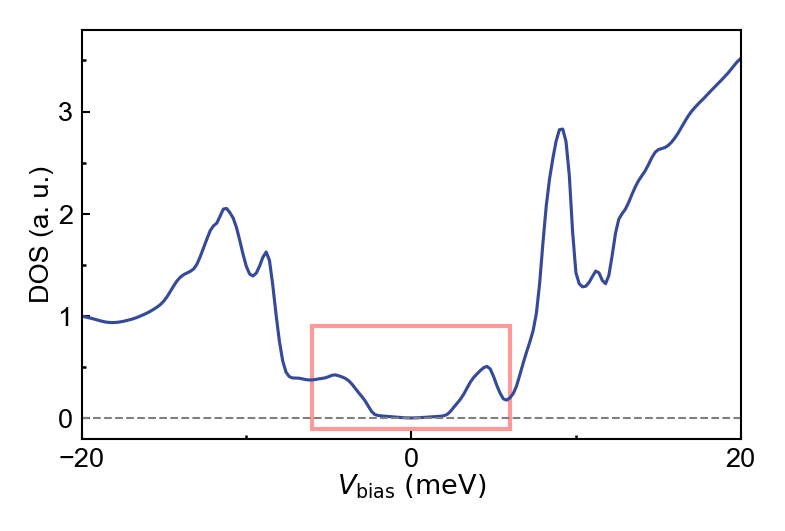}
\caption{\textbf{STS spectrum shown in the main text, displayed here on an expanded scale.} The red box marks the region presented in the main text. Additional resonance peaks appear at energies above the superconducting gap and disappear at temperatures exceeding $T_c$ \cite{yan2026stoichiometric}. The origin of these peaks and their relation to superconductivity are unclear.}
\label{fig:largeSTM}
\end{figure*}

\sisection{Details of theoretical modeling and additional calculations}\label{sec:theory}

\sisubsection{Details of the single-particle model}
The single-particle Hamiltonian $H_0$ used in the low-energy model is derived from Wannierization of low-energy density functional theory (DFT) bands, as discussed in Methods.
The details of $H_0$ and its derivation were given in Ref.~\cite{yan2026stoichiometric} and its supporting material. Here, we reproduce the relevant details for completeness.

Using the band structure of a DFT calculation without spin-orbit coupling, we perform Wannierization for bands within $\pm 1$~eV of the Fermi level at $k_z=0$, using Fe $d$ orbitals as trial states.
The resulting tight-binding Hamiltonian hosts two Fe atoms per unit cell (UC), the same as bulk FeTe.
In the Fe plane, Te atoms have a small effect on the hopping parameters between Fe atoms ~\cite{eschrig2009tbmodel,wu2016topological}. Therefore, the lattice can be further unfolded in the 2D Fe-plane, forming a 1-Fe lattice with primitive lattice vectors
$\mathbf{a}_\text{Fe,1}=\frac{1}{2}\mathbf{a}_\text{FeTe,1}-\frac{1}{2}\mathbf{a}_\text{FeTe,2}$, and $\mathbf{a}_\text{Fe,2}=\frac{1}{2}\mathbf{a}_\text{FeTe,1}+\frac{1}{2}\mathbf{a}_\text{FeTe,2}$, where $\mathbf{a}_\text{FeTe,1}$ and $\mathbf{a}_\text{FeTe,2}$ are the primitive lattice vectors of FeTe in the Fe plane. This is shown in Fig.~5(c) in the main text.
In this work, we always refer to the 1-Fe UC lattice constant and bases by $a_\text{Fe}$, $\mathbf{a}_\text{Fe,1}$, and $\mathbf{a}_\text{Fe,2}$, respectively, while $a\equiv a_\text{FeTe}$, $\mathbf{a}_1\equiv\mathbf{a}_\text{FeTe,1}$, and $\mathbf{a}_2\equiv\mathbf{a}_\text{FeTe,2}$ are reserved for the 2-Fe UC.
In addition, the high symmetry momentum points in the 1-Fe Brillouin zone (BZ) will be denoted with a prime, i.e., $\Gamma'$, $M'$, $X'$, and $Y'$, to distinguish them from those of the 2-Fe BZ.
Going from the 1-Fe lattice to the 2-Fe lattice doubles the UC, resulting in the BZ folding in the following manner: $M'\to\Gamma$, and $X'$ and $Y'\to M$.

\begin{figure}
\includegraphics[width=0.95\linewidth]{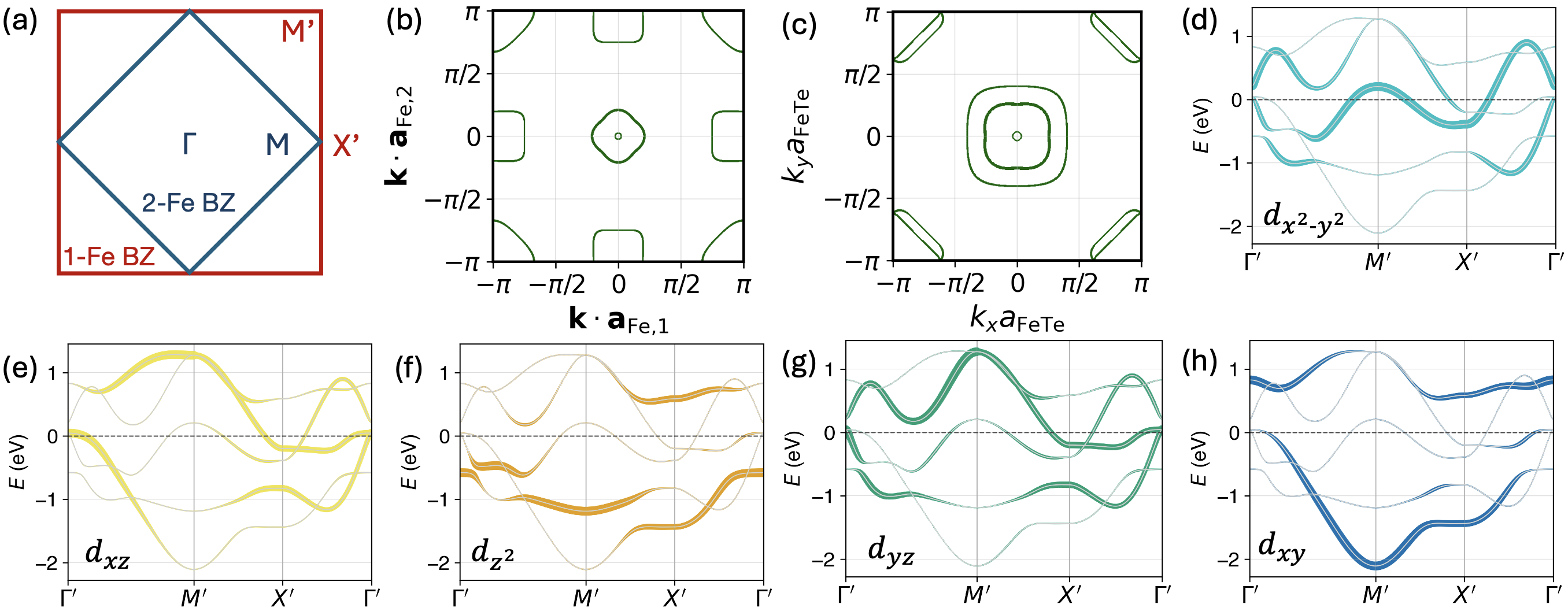}
\caption{
\textbf{Single-particle Fermi surface and band structure.}
(a) First Brillouin zone (BZ) of the 2-Fe UC and the 1-Fe UC.
(b) Fermi surface of the single-particle Hamiltonian $H_0$ in the 1-Fe BZ.
(c) Fermi surface of $H_0$ in the 2-Fe BZ.
(d)--(h) Single-particle band structure of $H_0$ in 1-Fe BZ and its orbital content among the Fe $d$ orbitals.
Note that the quantization axes of the atomic orbitals are along primitive lattice vectors of the FeTe UC.
}
\label{fig:app:single-particle}
\end{figure}

The unfolded Hamiltonian is further truncated so that the maximal hopping distance is $2\sqrt{2}a_\text{Fe}$.
This results in the truncated single-particle Hamiltonian $H_0$, which still accurately reproduces the DFT dispersion.
It takes the form
\begin{equation}
H_0 = \sum_{\mathbf{R},\mathbf{R}',\alpha,\beta,\sigma}^{|\mathbf{R}-\mathbf{R}'|\le2\sqrt{2}a_\text{Fe}} t^{\alpha\beta}_{\mathbf{R}-\mathbf{R}'} c^{\dagger}_{\mathbf{R}\alpha\sigma}c_{\mathbf{R}'\beta\sigma},
\end{equation}
where $\mathbf{R}=l_1 \mathbf{a}_\text{Fe,1}+l_2 \mathbf{a}_\text{Fe,2}$ is the lattice vector, $\sigma=\uparrow,\downarrow$ denotes the spin, while $\alpha$ and $\beta$ run over the Fe $d$ orbitals: $d_{z^2}, d_{xz}, d_{yz}, d_{x^2-y^2}$, and $d_{xy}$. Note that the quantization axes of the atomic orbitals are along primitive lattice vectors of the FeTe UC. The hopping matrices $t_{\mathbf{R}-\mathbf{R}'}$ used are taken from Ref.~\cite{yan2026stoichiometric}, and are presented below in units of eV:
\begin{equation}
t_{\mathbf{0}}= \operatorname{diag}(5.9860,6.6040,6.6040,6.3940,6.0470),
\end{equation}
\begin{equation}
t_{\mathbf{a}_\text{Fe,2}}=
\begin{pmatrix}
 0.0969 & -0.0760 &  0.0760 &  0      &  0.2738 \\
 0.0760 & -0.1475 &  0.0804 &  0.1579 &  0.1843 \\
-0.0760 &  0.0804 & -0.1475 &  0.1579 & -0.1843 \\
 0      & -0.1579 & -0.1579 & -0.0012 &  0      \\
 0.2738 & -0.1843 &  0.1843 &  0      &  0.3429
\end{pmatrix},
\end{equation}
\begin{equation}
t_{\mathbf{a}_\text{Fe,1}+\mathbf{a}_\text{Fe,2}}=
\begin{pmatrix}
-0.0807 &  0      & -0.1424 &  0.0067 &  0      \\
 0      &  0.0755 &  0      &  0      & -0.0654 \\
 0.1424 &  0      &  0.2179 &  0.1045 &  0      \\
 0.0067 &  0      & -0.1045 &  0.0748 &  0      \\
 0      &  0.0654 &  0      &  0      & -0.0034
\end{pmatrix},
\end{equation}
\begin{equation}
t_{2\mathbf{a}_\text{Fe,2}}=
\begin{pmatrix}
 0.0164 & -0.0241 &  0.0241 &  0      &  0.0350 \\
 0.0241 &  0.0015 & -0.0046 &  0.0168 & -0.0155 \\
-0.0241 & -0.0046 &  0.0015 &  0.0168 &  0.0155 \\
 0      & -0.0168 & -0.0168 &  0.0510 &  0      \\
 0.0350 &  0.0155 & -0.0155 &  0      & -0.0408
\end{pmatrix},
\end{equation}
\begin{equation}
t_{\mathbf{a}_\text{Fe,1}+2\mathbf{a}_\text{Fe,2}}=
\begin{pmatrix}
-0.0104 &  0.0046 & -0.0190 & -0.0004 &  0.0125 \\
-0.0046 &  0.0105 &  0.0010 & -0.0104 & -0.0101 \\
 0.0190 &  0.0010 & -0.0168 &  0.0474 & -0.0369 \\
-0.0004 &  0.0104 & -0.0474 &  0.0011 & -0.0227 \\
 0.0125 &  0.0101 &  0.0369 & -0.0227 &  0.0123
\end{pmatrix},
\end{equation}
\begin{equation}
t_{2\mathbf{a}_\text{Fe,1}+2\mathbf{a}_\text{Fe,2}}=
\begin{pmatrix}
-0.0114 &  0      & -0.0248 & -0.0116 &  0      \\
 0      & -0.0111 &  0      &  0      &  0.0066 \\
 0.0248 &  0      &  0.0267 &  0.0123 &  0      \\
-0.0116 &  0      & -0.0123 & -0.0299 &  0      \\
 0      & -0.0066 &  0      &  0      &  0.0144
\end{pmatrix}.
\end{equation}
The remaining hopping parameters can be obtained with inversion and
mirror-symmetry operations. These include reflections about the planes perpendicular
to $\mathbf{a}_\text{Fe,1}$, $\mathbf{a}_\text{Fe,2}$, and $\mathbf{a}_\text{Fe,1}-\mathbf{a}_\text{Fe,2}$.
We note that the mirror operations about the planes perpendicular to $\mathbf{a}_\text{Fe,1}$ and $\mathbf{a}_\text{Fe,2}$ are
effective symmetries of the 2D low-energy model. They are violated by the presence of Te atoms in FeTe.
A chemical potential term is further included and adjusted to ensure 6 electrons per Fe.
The resulting noninteracting Fermi surface and band structure are plotted in the 1-Fe BZ in Fig.~\ref{fig:app:single-particle}.
As one can see, the electron pockets at $X'$ and $Y'$ in the 1-Fe BZ ($M$ pockets in the 2-Fe BZ) derive mainly from the $d_{x^2-y^2}$ and $d_{xz}$ orbitals. The hole pocket at $M'$ (outermost pocket at $\Gamma$) is predominantly contributed by $d_{x^2-y^2}$.
Lastly, the two hole pockets at $\Gamma'$ (inner pockets at $\Gamma$) are mainly contributed by $d_{xz}$ and $d_{yz}$. In the presence of superconductivity, these $\Gamma'$ pockets are pushed below the Fermi level and disappear, as further discussed below.

\sisubsection{Interaction and Fermi surface renormalization}

\begin{figure}
\includegraphics[width=0.9\linewidth]{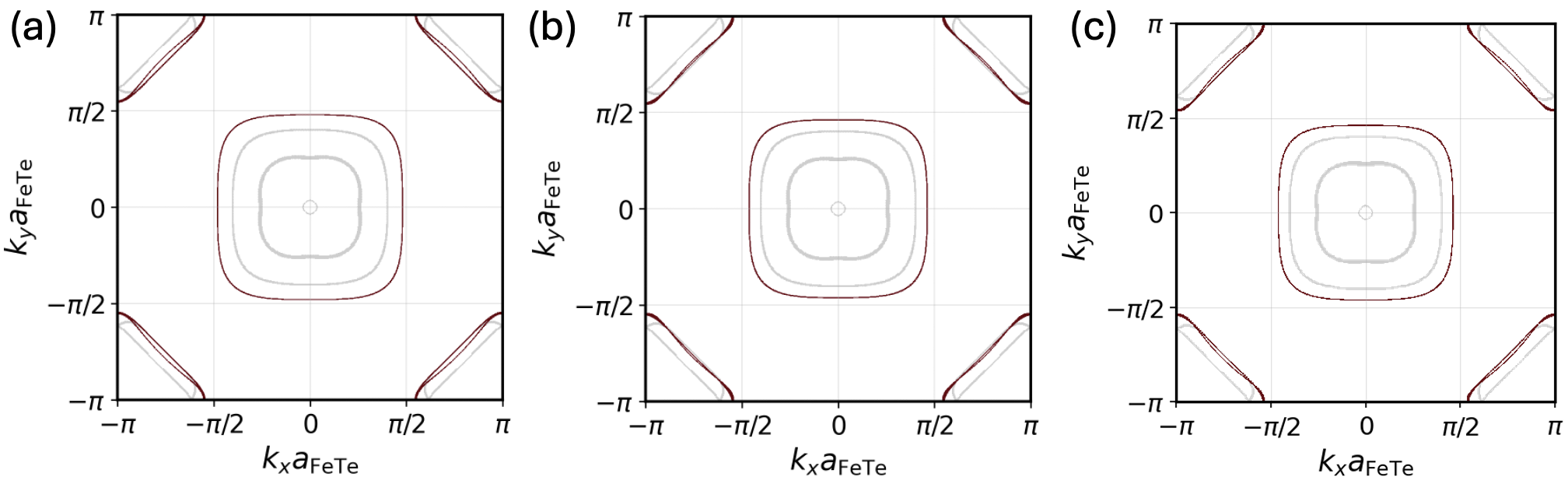}
\caption{
\textbf{Renormalized Fermi surface of the normal block in the BdG Hamiltonian in the interacting model.}
The interaction parameters used are $U'=U-2J$, $J=\frac{U}{4}$, $J'=J$, and for
(a) the nodal $d$-wave pairing solution presented in the main text:
$U=262.5$~meV,
$J_3=30$~meV,
$V_2=200$~meV,
$V_3=300$~meV;
for (b) the alternative nodal $d$-wave pairing:
$U=262.5$~meV,
$J_3=26.25$~meV,
$V_2=325$~meV,
$V_3=112.5$~meV;
for (c) the nodal $s$-wave pairing:
$U=375$~meV,
$J_3=37.5$~meV,
$V_2=200$~meV,
$V_3=150$~meV,
$V_4=100$~meV.
For comparison, the single-particle Fermi surface is plotted in gray.
The Fermi sea volume increased by (a) 0.019, (b) 0.036, and (c) 0.045.
}
\label{fig:app:interacting-FS}
\end{figure}

As discussed in the Methods, the Hubbard-Kanamori onsite repulsion $H_\text{int,onsite}$ \cite{gastiasoro2016unconventional,brydon2011magnetic,wu2016topological} is added to the single particle Hamiltonian following Ref.~\cite{yan2026stoichiometric}.
It includes onsite Hubbard repulsion $U$ between opposite spin electrons, the onsite Coulomb repulsion $U'$ between electrons in different orbitals, the Hund's coupling $J$, and the pair-hopping energy $J'$.
In addition, a next-next-nearest neighbor (NNNN) antiferromagnetic interaction $H_{\text{int},J_3}$~\cite{ma2009firstprinciples,Ducatman2014} between Fe atoms at $\pm2\mathbf{a}_\text{Fe,1}$ and $\pm2\mathbf{a}_\text{Fe,2}$ apart, with magnitude $J_3$, is also included to capture the double-striped antiferromagnetic order in as-grown FeTe samples with interstitial Fe~\cite{Rodriguez2011,yan2026stoichiometric}.

To achieve superconductivity, we include attractive interactions $H_\text{SC}$. The commonly adopted superconducting pairing in FeTe is the nonlocal  pairing to the four next-nearest-neighbor (NNN) Fe atoms at $\pm\mathbf{a}_\text{Fe,1}\!\pm\mathbf{a}_\text{Fe,2}$.
At the single-particle level, the pairing amplitude between NNN sites takes the form $\cos(\mathbf{k}\cdot \mathbf{a}_\text{Fe,1})\cos(\mathbf{k}\cdot \mathbf{a}_\text{Fe,2})$ in the momentum space of the 1-Fe BZ, which becomes zero only when $\mathbf{k}\cdot \mathbf{a}_\text{Fe,1} =\pm\pi/2$ or $\mathbf{k}\cdot \mathbf{a}_\text{Fe,2}=\pm\pi/2$.
Adding mean-field NNN pairing directly to our single-particle model results in a fully gapped $s_\pm$-wave superconducting state with pairing amplitudes of one sign on the 1-Fe $X'$ and $Y'$ pockets (2-Fe $M$ pockets) and the opposite sign on the 1-Fe $M'$ pocket (2-Fe outermost $\Gamma$ pocket).
Therefore, the NNN pairing itself cannot capture the two-gap behavior observed in experiments with the V-shaped inner gap in the local density of states (DOS), which is characteristic of a pairing node.
With these insights from the experiment, we consider both the attractive NNN interaction with strength $V_3$ and the attractive nearest-neighbor (NN) interaction between Fe atoms $\pm\mathbf{a}_\text{Fe,1}$ and $\pm\mathbf{a}_\text{Fe,2}$ apart, with strength $V_2$.
The mean-field solution of these parameters converges to nodal $d$-wave spin-singlet pairing states with DOS line shapes that resemble the STS measurement.
The nodal pairing can also be realized with $s$-wave pairing.
To study the nodal $s$-wave pairing, we further incorporate the attractive NNNN interaction with strength $V_4$, between Fe atoms $\pm2\mathbf{a}_\text{Fe,1}$ and $\pm2\mathbf{a}_\text{Fe,2}$ apart.

Now we discuss the symmetries present in our model. The full interacting Hamiltonian is solved self-consistently within the mean-field (MF) approximation, under the assumption that (i)~the system remains translationally invariant, and (ii)~there is no spin flipping, i.e., $S_z$ is a good quantum number.
The model further preserves the spin-$\mathrm{SU}(2)$ symmetry, and
the spinful time-reversal symmetry.
They are enforced in the self-consistent solver.
The MF Bogoliubov-de-Gennes (BdG) Hamiltonian also hosts the intrinsic particle-hole symmetry.
Combining the time-reversal symmetry, spin $\mathrm{SU}(2)$ symmetry and the particle-hole symmetry leads to the spinful and spinless chiral symmetries in our model.
They allow us to use the chiral winding number to topologically identify the nodal points, as discussed in the main text and in Methods.
The point group symmetries are also present in the system, including inversion and mirrors normal to $\mathbf{a}_\text{FeTe,1}+\mathbf{a}_\text{FeTe,2}$, $\mathbf{a}_\text{FeTe,1}-\mathbf{a}_\text{FeTe,2}$ and $\mathbf{a}_\text{FeTe,1}$ directions.
Unlike the case of $\mathrm{SU}(2)$ and the time-reversal symmetries, the spatial symmetries are not explicitly enforced and can be broken in the converged solutions.
In particular, the mirror symmetry normal to $\mathbf{a}_\text{FeTe,1}$ is violated by the anomalous block, i.e., the particle-hole sector, of the mean-field Hamiltonian for both of the nodal $d$-wave solutions presented.
For the $s$-wave solution, all spatial symmetries remain intact.

During the self-consistent iterations, the solver adjusts the chemical potential to keep the system at 6 electrons per Fe.
For each set of the interacting parameters, we initialize the self-consistent solver with different MF order parameter sets, including fully-random, superconducting, and spin-imbalanced cases.
The MF solver runs on an $L\times L$ translationally invariant lattice. The converged MF parameters and the associated MF BdG Hamiltonian can then be used to compute observables for lattices of any desired size.
For the BdG pairing distribution plots, we used a lattice of $L=1200$.
For the DOS plots, we used a hybrid momentum mesh. For $d$-wave pairing cases, a coarse momentum mesh with $L=240$ is first used to sample the full BZ. Then, for momenta within $\pm25$~meV of the Fermi level, we subdivide the momentum mesh plaquette into a $100^2$-times finer grid. This generates a smooth electron DOS trace computed by $\frac{\mathrm{d}I}{\mathrm{d}V} \propto \sum_{n}\int \mathrm{d}^2\mathbf{k} \|u_{n\mathbf{k}}\|^2 \frac{\eta}{\pi}
[ \left(eV_{\text{bias}}-\varepsilon_{n\mathbf{k}}\right)^2+\eta^2
]^{-1}$, where $\varepsilon_{n\mathbf{k}}$ is the $n$th eigenenergy at momentum $\mathbf{k}$, $u_{n\mathbf{k}}$ is the particle component of the eigenstate $(u_{n\mathbf{k}},v_{n\mathbf{k}})^T$, and $\eta=0.05$~meV is the broadening.
For the $s$-wave case, we employed a higher momentum resolution to produce the DOS data. The coarse momentum mesh is $L=1200$, with a $100^2$ subdivision for momenta within $\pm10$~meV of the normal-block Fermi surface, and a broadening of 0.01~meV.

By tuning the strengths $U$, $J_3$, $V_2$, and $V_3$, we obtain the nodal $d$-wave superconductivity on the $\Gamma$ pocket for two sets of interaction parameters.
The first set of the interaction parameters is
$U=262.5$~meV,
$J_3=30$~meV,
$V_2=200$~meV,
$V_3=300$~meV,
$U'=U-2J$, $J=\frac{U}{4}$, and $J'=J$.
The MF parameters converged on a lattice of $L=24$.
They are used to generate the results discussed in the main text and Methods.
In the main text, Figure 5 shows the superconducting gap in our theoretical model on the Fermi surface, as well as the DOS of the superconducting state at zero temperature. They show good agreement with the experimental results.
As discussed in the main text, the observed two-gap behavior can be captured by another set of interaction parameters with $U=262.5$~meV,
$J_3=26.25$~meV,
$V_2=325$~meV,
$V_3=112.5$~meV,
$U'=U-2J$, $J=\frac{U}{4}$, and $J'=J$.
In addition, the nodal two-gap superconductivity can also be realized under $s$-wave pairing when incorporating $V_4$ for NNNN attractive interaction.
For this case, we used $U=375$~meV,
$J_3=37.5$~meV,
$V_2=200$~meV,
$V_3=150$~meV,
$V_4=100$~meV,
$U'=U-2J$, $J=\frac{U}{4}$, and $J'=J$.
Further discussion of the results for each of these parameter sets is presented in the next section. We emphasize that the pairing nodes are robust to small changes of interaction parameters as they are protected by the chiral symmetry.

An important consequence of the interaction is the renormalization of the normal block of the MF Hamiltonian, i.e., its particle-particle sectors.
The renormalization leads to a significant change of the Fermi surface,  where the BdG gap opens.
In Fig.~\ref{fig:app:interacting-FS}, we show the Fermi surface and band structure in the normal block with only the particle-particle terms of the MF BdG Hamiltonian under the three sets of interaction parameters used.
Compared to the single-particle Fermi surface, we find that the $\Gamma'$ pockets in the 1-Fe BZ (inner pockets at $\Gamma$ in 2-Fe BZ) are suppressed. Meanwhile, all the other Fermi pockets are enlarged, i.e., the electron pockets at $M$ and the outer hole pocket at $\Gamma$.
Because this Fermi surface is an auxiliary surface extracted from the normal block of a self-consistent superconducting BdG Hamiltonian, its enclosed area and Luttinger count are not constrained to reproduce the fixed electron density. This is reflected in the increase in the volume of the Fermi sea by 0.019 in Fig.~\ref{fig:app:interacting-FS}(a), by 0.036 in Fig.~\ref{fig:app:interacting-FS}(b), and by 0.045 in Fig.~\ref{fig:app:interacting-FS}(c).
We also confirmed that no other order exists in the Nambu-diagonal sectors in the converged MF solution, i.e., in the particle-particle and hole-hole blocks, such as spatial symmetry breaking or magnetism.
The MF ground state is thus superconducting, with the BdG gap and nodal points sitting on the renormalized Fermi surface.

\sisubsection{Pocket-resolved contribution to the DOS}

\begin{figure}
\includegraphics[width=0.7\linewidth]{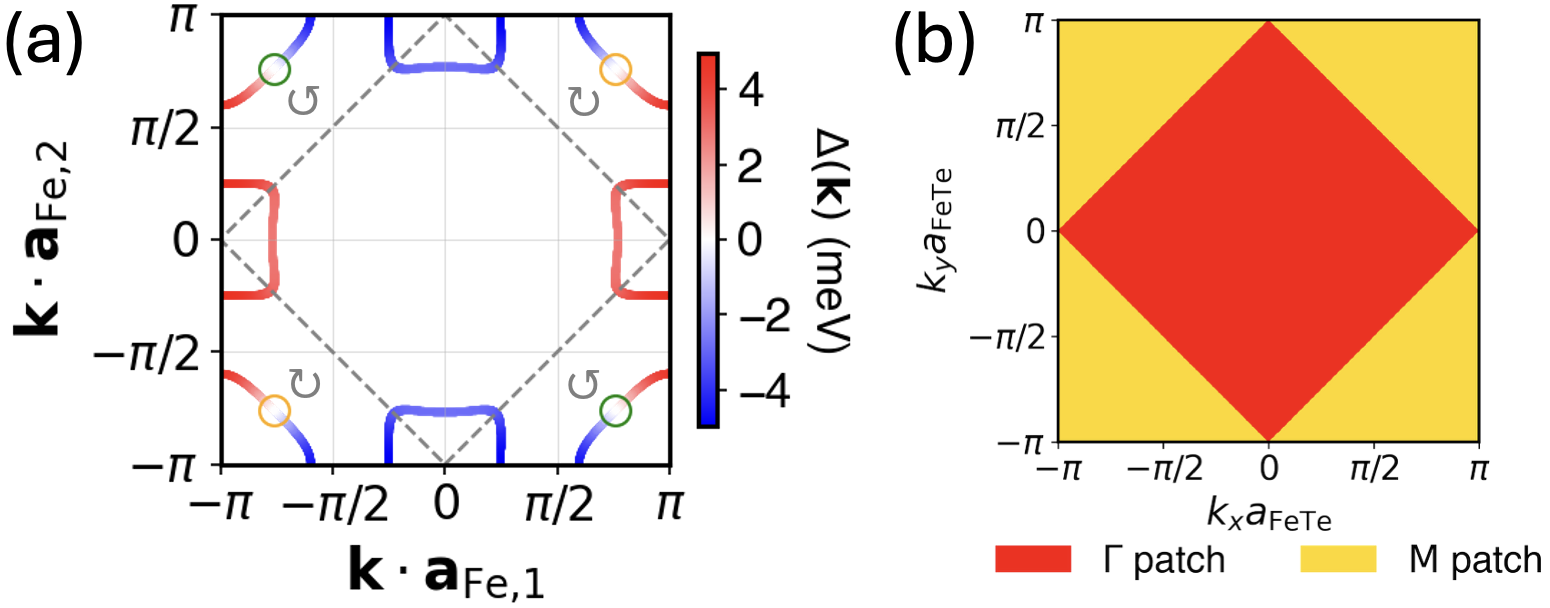}
\caption{
\textbf{Supplements to Fig.~5 in the main text.}
(a)~Projected pairing amplitude $\Delta(\mathbf{k})$ on the interaction-renormalized Fermi surface in the 1-Fe Brillouin zone (BZ) for the $d$-wave state presented in the main text, corresponding to the unfolded representation of Fig.~5(d). Nodal points are labeled by their chiral windings: green and yellow circles indicate winding numbers of $+1$ and $-1$, respectively. Dashed lines mark the boundary of the 2-Fe BZ. The two $M$-centered pockets in the 2-Fe Brillouin zone originate from the $X'=(\pi,0)$ and $Y'=(0,\pi)$ electron pockets of the unfolded 1-Fe Brillouin zone. Upon folding, $X'$ and $Y'$ map onto the same $M$ point while retaining their distinct band characters. The anisotropic pairing amplitudes on the $X'$ and $Y'$ pockets are correspondingly mapped onto the two $M$-centered pockets, giving gap magnitudes ranging from a minimum of $\sim3$~meV to a maximum of $\sim5$~meV.
(b)~Momentum patches used to resolve the DOS contributions from the Fermi pockets at $\Gamma$ and $M$.
}

\label{fig:app:suppFig5}
\end{figure}

Here we discuss the details of the superconducting gap and the DOS contributions in the theoretical model.
The experimental STS result gives the most direct guidance for our parameter selection.
We briefly recall that for the DOS in an effectively 2D superconductor,
the local minimum of the BdG gap on a connected component of the Fermi surface, which is also a local minimum in the 2D BZ, shows up as a step in the DOS.
On the other hand, the local maximum of the BdG gap on a connected component of the Fermi surface is a saddle point in the 2D BZ and shows up as a logarithmically diverging peak in the DOS.
Lastly, a nodal point shows up as a V-shaped gap in the DOS.

In the STS trace shown in Fig.~5, a V-shaped gap is observed around zero bias, accompanied by an upturn in the DOS at approximately $2$~mV and a peak at $4.6$~mV, together with their corresponding features at negative bias.
Before the upturn of the V-shaped gap, there may be a weak shoulder-like feature that is difficult to resolve unambiguously.
Depending on the interpretation of this feature, we identify two sets of interaction parameters in our theoretical model whose MF solutions are consistent with the experimental data.

The first is the nodal $d$-wave pairing presented in the main text and the Methods. Its interaction parameters are
$U=262.5$~meV,
$J_3=30$~meV,
$V_2=200$~meV,
$V_3=300$~meV,
$U'=U-2J$, $J=\frac{U}{4}$ and $J'=J$, and the MF solution converges on an $L=24$ lattice.
Figure 5(d) in the main text shows its projected pairing distribution in the 2-Fe BZ, while Fig.~\ref{fig:app:suppFig5}(a) shows the same data in the 1-Fe BZ to better distinguish the two pockets at $M$, i.e., $X'$ and $Y'$.
The total DOS together with the contributions from the $\Gamma$ and $M$ pockets are plotted in Fig.~5(e) in the main text, using the patches of the first BZ depicted in Fig.~\ref{fig:app:suppFig5}(b).
At the $\Gamma$ pocket, the minima of the BdG gap are zero due to the nodal points, which contribute to the V-shaped inner gap in DOS. These nodal points carry chiral winding numbers of $\pm1$ in an alternating manner around the pocket at intersections $k_x=0$ and $k_y=0$ of the Fermi surface. Note that mirror symmetries relate nodes at image momenta with opposite chiral windings. Therefore, the mirror symmetry normal to $\mathbf{a}_\text{FeTe,1}$ has to be broken to allow nodal points on the $k_x=0$ and $k_y=0$ lines.
The maxima on the $\Gamma$ pocket at 9.04~meV contribute to the DOS peak at $V_\text{bias}=4.52$~mV.
On the two $M$ pockets, the minimum of the BdG gap is 5.63~meV, setting the width of the U-shaped gap and contributing to the step in DOS at $V_\text{bias}=2.81$~mV, while the maxima at 9.98~meV contribute to the DOS peak at 4.99~mV in Fig.~5(e).
Together, these features clearly demonstrate the two-gap behavior.

\begin{figure}
\includegraphics[width=1\linewidth]{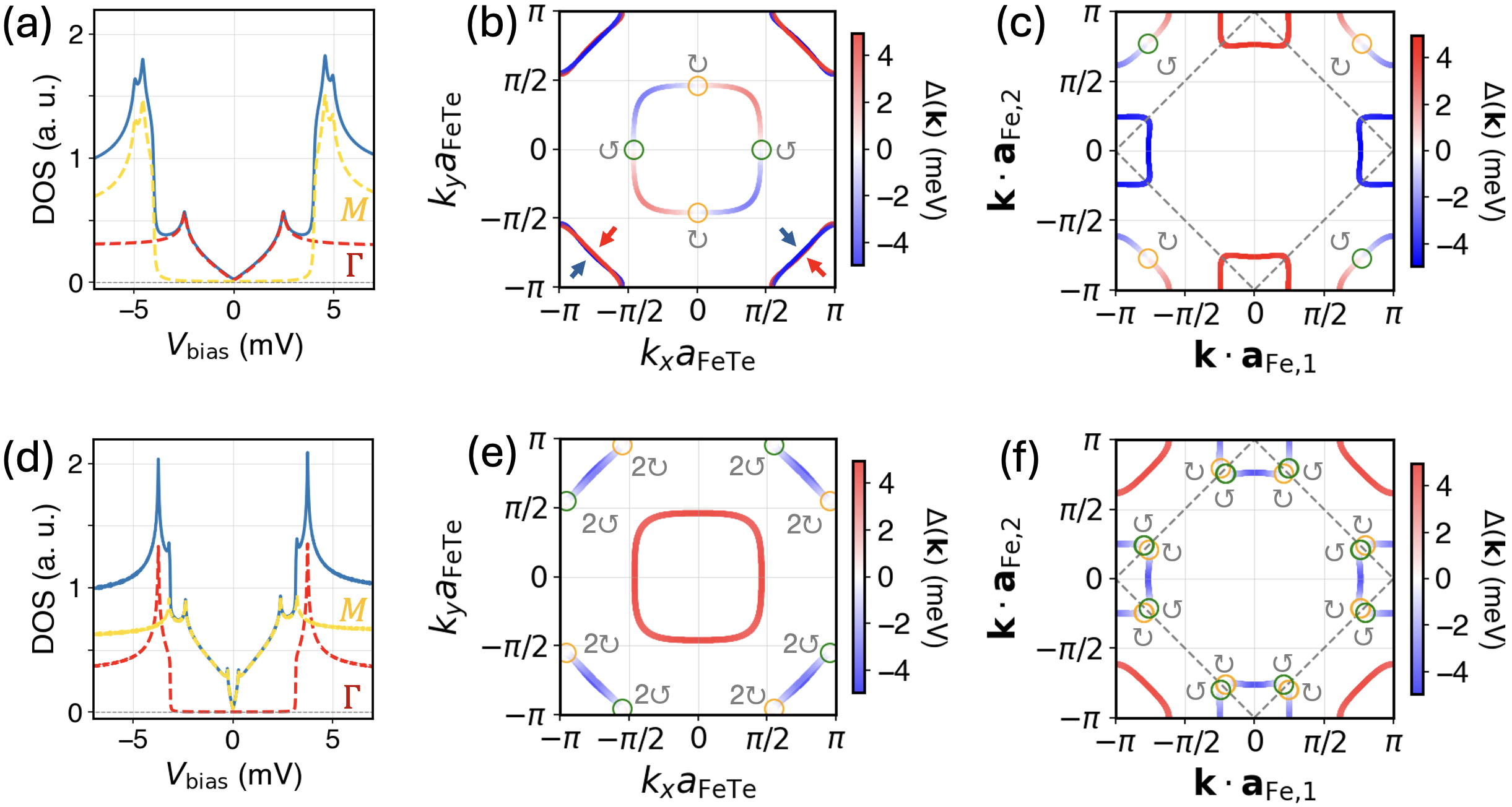}
\caption{
\textbf{DOS and the projected BdG pairing of alternative interacting MF solutions.}
(a--c)~\emph{Results of the alternative $d$-wave pairing setup}, using interaction parameters
$U=262.5$~meV,
$J_3=26.25$~meV,
$V_2=325$~meV,
$V_3=112.5$~meV,
$U'=U-2J$, $J=\frac{U}{4}$, and $J'=J$.
(a)~DOS of the MF BdG Hamiltonian, with Fermi-pocket-resolved contributions.
(b)~Projected pairing $\Delta(\mathbf{k})$ on the interaction-renormalized Fermi surface in the 2-Fe BZ, where nodal points are labeled by their chiral windings. Green circles indicate a $+1$ chiral winding number, while yellow circles indicate $-1$. Arrows indicate the two $M$ pockets with opposite signs of pairing.
(c)~$\Delta(\mathbf{k})$ in the 1-Fe BZ. Dashed lines indicate the 2-Fe BZ.
(d--f)~\emph{Results of the $s$-wave pairing setup}, using interaction parameters
$U=375$~meV,
$J_3=37.5$~meV,
$V_2=200$~meV,
$V_3=150$~meV,
$V_4=100$~meV,
$U'=U-2J$, $J=\frac{U}{4}$, and $J'=J$.
(d)~DOS of the MF BdG Hamiltonian, with Fermi-pocket-resolved components.
(e)~$\Delta(\mathbf{k})$ on the interaction-renormalized Fermi surface in the 2-Fe BZ, where nodal points are labeled by their chiral windings.
Each displayed nodal location contains two closely spaced nodes with the same chiral winding number, originating from the $X'$ and $Y'$ pockets of the 1-Fe BZ; their separation is too small to be resolved on the plotted scale.
(f)~$\Delta(\mathbf{k})$ on the interaction-renormalized Fermi surface in the 1-Fe BZ, obtained by unfolding the 2-Fe BZ in (e).
}

\label{fig:app:dos}
\end{figure}

To match the possibility that the shoulder comes from a secondary peak in the DOS,
we identify another set of interacting parameters that generates nodal $d$-wave pairing:
$U=262.5$~meV,
$J_3=26.25$~meV,
$V_2=325$~meV,
$V_3=112.5$~meV,
$U'=U-2J$, $J=\frac{U}{4}$, and $J'=J$, and the MF solution converges on an $L=24$ lattice.
Figure~\ref{fig:app:dos}(b) and (c) show its gap distribution on the interaction-renormalized Fermi surface, in the 2-Fe BZ and 1-Fe BZ, respectively.
Figure \ref{fig:app:dos}(a) shows the total DOS together with the contributions from the $\Gamma$ and $M$ pockets.
It also hosts nodal points on the $\Gamma$ pocket in the same manner as in the previous $d$-wave-pairing case.
In this case, the maximum BdG gap in the $\Gamma$ pocket is 4.94~meV, which contributes to the DOS peak at 2.48~mV and coincides with the possible shoulder feature in the experimental data.
On $M$ pockets, the minimum of the BdG gap is 8.01~meV, contributing to the DOS step at 4~mV, while the local maxima of 9.11~meV and 9.91~meV contribute to the DOS peaks at 4.56~mV and 4.94~mV, respectively.
The effect of pairing is limited to within $V_\text{bias}<10$~mV, as indicated by the small polarization toward pairing with $P_\text{SC}=0.14$ at 10~mV.

Finally, we present a nodal $s$-wave pairing state produced by the interacting model with NNNN attractive interaction included. The interaction parameters are
$U=375$~meV,
$J_3=37.5$~meV,
$V_2=200$~meV,
$V_3=150$~meV,
$V_4=100$~meV,
$U'=U-2J$, $J=\frac{U}{4}$, and $J'=J$.
The nodal points here are located on the $M$ pockets, as shown in Fig.~\ref{fig:app:dos}(e) and (f) in the 2-Fe BZ and 1-Fe BZ, respectively. Since the converged mean-field solution preserves the mirror symmetry normal to $\mathbf{a}_\text{FeTe,1}$, the nodal points come in image pairs of opposite chiral winding numbers about all mirror planes and cannot appear on those planes. This can be seen from the definition of the chiral winding number along a path $C$, as defined in Eq.~(6) in Methods, $W[C]=\frac{i}{4\pi}\oint_C \operatorname{Tr} [\mathcal{C} h^{-1}_\text{BdG}(\mathbf{k}) \mathrm{d} h_\text{BdG}(\mathbf{k}) ]$. Here, $\mathcal{C}$ is the chiral symmetry operator, represented by the Pauli-$y$ matrix acting on the particle-hole index. Since $\mathcal{C}$ commutes with spatial symmetries, a winding number integral at $\mathbf{k}$ is mapped to an integral at $-\mathbf{k}$ with opposite orientation. Consequently, chiral winding numbers of mirror-symmetry-related momenta must be opposite. On the other hand, inversion-symmetry-related momenta have the same chiral winding number, since inversion preserves the orientation of a closed path.

In the $s$-wave pairing case, there are in total 16 nodal points, 8 on each of the two $M$ pockets. Nodes on the same $M$ pocket are related by symmetry. They appear in pairs close to the intersections of the $k_x=\pi$ and $k_y=\pi$ lines with the $M$ pocket.
The maxima of the pairing gap in the $M$ pockets contribute to the peaks in the DOS in Fig.~\ref{fig:app:dos}(d) at 2.38~mV and 3.20~mV, respectively.
For the $\Gamma$ pocket, the gap minima correspond to the DOS step at $3.14$~mV, while the maxima correspond to the DOS peak at 3.74~mV.
The effect of pairing is also limited to within $V_\text{bias}<10$~mV, as indicated by the small polarization toward pairing with $P_\text{SC}=0.055$ at 10~mV.

\end{document}